\documentclass[
    amsmath,
    nofootinbib,
    amssymb,
    english,
    aps,
    showkeys,
]{revtex4-2}
\usepackage{fix-cm}
\usepackage[utf8]{inputenc}
\usepackage[english]{babel}
\usepackage{graphicx}
\usepackage{latexsym,amsthm,amsfonts}
\usepackage{makeidx}   
\usepackage[T1]{fontenc}
\usepackage{subfigure}
\usepackage{dcolumn}
\usepackage{bm}
\usepackage{cancel}
\usepackage{xcolor}
\usepackage[hyperindex,colorlinks,linkcolor=blue,citecolor=red]{hyperref}

\usepackage{textgreek}
\usepackage{comment}

\allowdisplaybreaks[1]

\usepackage{upgreek}
\usepackage{graphicx}
\usepackage{tensor}

\begin{document}
\title{Magnetically charged black-bounce solution via nonlinear electrodynamics in a k-essence theory}

\author{Carlos F. S. Pereira}
	\email{carlosfisica32@gmail.com}
	\affiliation{Departamento de Física, Universidade Federal do Espírito Santo, Av. Fernando Ferrari, 514, Goiabeiras, 29060-900, Vit\'oria, ES, Brazil.}
	
	\author{Denis C. Rodrigues}
    \email[]{deniscr@gmail.com}
	\affiliation{Núcleo Cosmo-ufes \& Departamento de Física, Universidade Federal do Espírito Santo, Av. Fernando Ferrari, 514, Goiabeiras, 29060-900, Vit\'oria, ES, Brazil.}

 \author{Marcos V. de S. Silva}
	\email{marcosvinicius@fisica.ufc.br}
	\affiliation{Departamento de Física, Programa de Pós-Graduação em Física,
Universidade Federal do Ceará, Campus Pici, 60440-900, Fortaleza, Ceará, Brazil}	
		
	\author{Júlio C. Fabris}
    \email[]{julio.fabris@cosmo-ufes.org}
	\affiliation{Núcleo Cosmo-ufes \& Departamento de Física, Universidade Federal do Espírito Santo, Av. Fernando Ferrari, 514, Goiabeiras, 29060-900, Vit\'oria, ES, Brazil}
    \affiliation{National Research Nuclear University MEPhI (Moscow Engineering Physics Institute), 115409, Kashirskoe shosse 31, Moscow, Russia.}
	
	\author{Manuel E. Rodrigues}
    \email[]{esialg@gmail.com}
	\affiliation{Faculdade de Ci\^encias Exatas e Tecnologia, Universidade Federal do Par\'a Campus Universit\'ario de Abaetetuba, 68440-000, Abaetetuba, Par\'a, Brazil and Faculdade de F\'isica, Programa de P\'os-Gradua\c{c}\~ao em F\'isica, Universidade Federal do Par\'a, 66075-110, Bel\'em, Par\'a, Brazil.}

    \author{H. Belich} 
    \email{humberto.belich@ufes.br}
    \affiliation{Departamento de F\'isica e Qu\'imica, Universidade Federal do Esp\'irito Santo, Av.Fernando Ferrari, 514, Goiabeiras, Vit\'oria, ES 29060-900, Brazil.}
	
\begin{abstract}
\noindent  In the present work, we obtain and analyze a new class of analytical solutions of magnetically charged black-bounces in the k-essence theory, spherically symmetric in (3+1)-dimensions, coupled to nonlinear electrodynamics (NED). We consider two metric models, Simpson-Visser and Bardeen-type black-bounces, for the k-essence configurations $n=1/3$ and $n=1/5$. We obtain in an analytical way the scalar field, the field potential, and Lagrangian NED, which are necessary to support the metrics. We analyze the behavior of these quantities and the energy conditions due to the scalar field and the NED.
\end{abstract}
	
	
\maketitle
	
\section{Introduction}\label{sec1}

General relativity in its classical form is the simplest theory to describe gravitational and cosmological scenarios. Several objects can be predicted within this framework, including black holes, wormholes, neutron stars, and interesting phenomena such as gravitational waves and light deviation in the presence of a strong field, among others \cite{INTRO1, INTRO2, INTRO3, INTRO4, INTRO5, LIGOScientific:2016aoc}. The best-known spherically symmetric black hole solution is the Schwarzschild solution. This solution is quite simple, as it is described solely by the mass of the black hole and has only one event horizon, which is a non-return surface, and one physical singularity, a point where geodesics are interrupted. Regular black holes emerge as alternatives that stand out compared to typical black holes due to the absence of singularities within their interiors, the first being proposed by Bardeen in 1968 \cite{INTRO6}. Later, Beato and Garcia showed that the matter content associated with the Bardeen metric that solved Einstein's equations was NED \cite{INTRO7}. Usually, static black holes with regularized centers can always have NED as their source of matter, the sources being either electric or magnetic \cite{Rodrigues:2018bdc, Bronnikov:2022ofk, Bronnikov:2024izh, Bolokhov:2024sdy}. The same is not necessarily true for rotating solutions \cite{Rodrigues:2017tfm}.

Recently, Simpson and Visser used the Schwarzschild metric with a regularization procedure, $r^2 \to r^2 + a^2$, where $a$ is the regularization parameter, to obtain the so-called black-bounce solutions \cite{matt}. This regularization method essentially removes the point where the singularity existed along with its surrounding neighborhood, thereby creating a sort of wormhole within a black hole. These spacetimes can be classified as two-way traversable wormholes ($a > 2m$), one-way wormholes ($a = 2m$), and regular black holes with symmetric horizons ($a < 2m$), where $a$ represents the throat of the wormhole. In the case $a = 2m$, the event horizon coincides with the position of the wormhole's throat, and in this situation, we have what is called a black throat. In works \cite{INTRO8, INTRO9}, it was verified that the matter content associated with the Simpson-Visser metric is a combination of NED and a phantom scalar field, where the electromagnetic source is a magnetic charge. It is important to note that these same solutions can also be obtained by considering an electrically charged source \cite{INTRO10}.

Following Simpson and Visser's work, a wide range of scenarios have been explored. These include modifications to area functions \cite{INTRO11}, modifications to the mass of the object that makes it depend on the radial coordinate \cite{MM2}, and a regularization of the Reissner–Nordström spacetime \cite{RN1, RN2}. Furthermore, there are works in the context of modified gravity \cite{INTRO12, INTRO13, INTRO14}, non-conservative theory \cite{INTRO15}, Braneworld \cite{INTRO16}, and black strings \cite{INTRO17, INTRO18, INTRO19}. This class of solutions has also been extended to spherically symmetric and stationary spacetimes \cite{RN1, INTRO21, INTRO18, INTRO23}, as well as to the study of light deflection and gravitational lensing effects \cite{INTRO24, INTRO25, INTRO26, INTRO27, INTRO28}. In the mass limit tending to zero, the Simpson-Visser spacetime transforms into the topologically charged Ellis-Bronnikov spacetime, and in this context quantum systems were explored \cite{INTRO29, INTRO30, INTRO31, INTRO32, INTRO33, INTRO34}.

Conventionally, in general relativity, solutions for regular black holes and traversable wormholes require that energy conditions be violated, thus requiring exotic matter. For the case of regular black holes with only NED, the strong energy condition is always violated. However, there are also solutions that violate other energy conditions \cite{Rodrigues:2017yry}. In the case of the scalar field, a canonical scalar field minimally coupled with gravitational theory is not sufficient to generate regular solutions, necessitating the presence of a phantom scalar field, which violates the null energy condition \cite{PRL}.

The initial proposal of this work was to investigate possible modifications in the energy conditions, for example, arising from NED incorporated into a k-essence theory. As we will see below, the electromagnetic sector is not modified by the presence of the k-essence function, however, the phantom and potential scalar field configurations are altered.

The paper is structured as follows: Section \ref{sec2} establishes the theoretical basis of the k-essence model in conjunction with NED, including the derivation of the equations of motion. In Sections \ref{sec3} and \ref{sec4}, we apply the methodology to two specific models and derive the physical quantities for the k-essence configurations of $n=1/3$ and $n=1/5$. In Section \ref{sec5}, we derive the energy conditions generically for each configuration, $n=1/3$ and $n=1/5$. In Section \ref{sec6}, we analyze the energy conditions for each model. Finally, the conclusions are presented in Section \ref{sec7}.

\section{General relations}\label{sec2}

The k-essence theories are characterized by the presence of scalar fields whose kinetic terms are introduced in a non-canonical way and have been recently explored in the context of wormhole solutions \cite{CDJM1, CDJM2}. Now, we will introduce a term that represents NED. Therefore, consider the model described by the action below:

\begin{equation}\label{Lagran}
S=\int{d^4}{x}\sqrt{-g}[R-F(X,\phi) + L(f)]\,,
\end{equation}
where $R$ is the Ricci scalar, $X=\eta\phi_{;\rho}\phi^{;\rho}$ denotes the kinetic term, and $L(f)$ represents the contribution from electromagnetism, where $f=\frac{H_{\mu\nu}H^{\mu\nu}}{4}$, with $H_{\mu\nu}=\partial_\mu{A_\nu}-\partial_\nu{A_\mu}$ being the electromagnetic tensor, and $A_\mu$ being the four-dimensional vector potential. Though k-essence models can include a potential term and non-trivial couplings, the scalar sector is typically minimally coupled to gravity. The parameter $\eta=\pm 1$ is introduced to avoid imaginary terms in the kinetic expression $X$. By selecting different forms for the function $F(X,\phi)$, k-essence theories can describe both phantom \cite{phan1, phan2, phan3, phan4} and standard scalar fields.

By varying the above action (\ref{Lagran}) with respect to the fields and the metric tensor, we obtain the following equations of motion:

\begin{eqnarray}\label{1}
 G_{\mu\nu}=T^{\phi}_{\mu\nu} + T^{EM}_{\mu\nu}, \\\label{2} 
\eta\nabla_\alpha\left(F_X\phi^{\alpha}\right)-\frac{1}{2}F_\phi=0, \\\label{3}
\nabla_\mu\left[L_{f}H^{\mu\nu}\right]=0,
\end{eqnarray} where $G_{\mu\nu}$ is the Einstein tensor, $T^{\phi}_{\mu\nu}$ and $T^{EM}_{\mu\nu}$ are the stress-energy tensors of the scalar field $\phi$ and the electromagnetic field, respectively, $F_X=\frac{\partial{F}}{\partial{X}}$, $F_\phi=\frac{\partial{F}}{\partial\phi}$, $\phi_\mu=\partial_\mu\phi$, and $L_f=\frac{\partial{L}}{\partial{f}}$.

The energy-momentum tensor for each of the fields is defined by:

\begin{eqnarray}\label{4}
T^{\phi}_{\mu\nu}= -\frac{F}{2}g_{\mu\nu} + \eta{F_{X}}\nabla_\mu{\phi}\nabla_\nu{\phi}, \\\label{5}
T^{EM}_{\mu\nu}= \frac{L(f)}{2}g_{\mu\nu} -\frac{L_f}{2}{H_\mu}^\alpha{H_{\nu\alpha}}.
\end{eqnarray}

The line element representing the most general spherically symmetric and static spacetime takes the form:
\begin{eqnarray}\label{6}
ds^2=e^{2\gamma\left(u\right)}dt^2-e^{2\alpha\left(u\right)}du^2-e^{2\beta\left(u\right)}d\Omega^2,
\end{eqnarray} where $u$ is an arbitrary radial coordinate and $d\Omega^2 = d\theta^2 + \sin^2\theta  d\varphi^2$ is the area element. Since our spacetime is spherically symmetric and static, we can assume that the scalar potential is a function only of the radial coordinate, $\phi = \phi\left(u\right)$.

For our purposes, we are only interested in possible magnetically charged solutions. Thus, the non-zero component of the Maxwell-Faraday tensor is given by

\begin{eqnarray}\label{7}
    H_{23}= q_{m}\sin\theta,
\end{eqnarray} where the electromagnetic scalar is defined by

\begin{eqnarray}\label{8}
    f(u)=\frac{q^2_m}{2e^{4\beta(u)}},
\end{eqnarray} with $q_m$ being the magnetic charge.

Thus, we write the general equations of motion, which are the same as those contained in Refs. \cite{CDJM1, CDJM2, KDJ}. However, they are now modified by NED. It is assumed that the function $X = -\eta e^{-2\alpha} (\phi')^2$ is positive, which implies that $\eta = -1$. As a result, the equations of motion take the form:

\begin{eqnarray}\label{9}
2\left(F_X{e^{-\alpha+2\beta+\gamma}}\phi'\right)' - {e^{\alpha+2\beta+\gamma}}F_\phi=0, \\\label{10}
{\gamma}'' + {\gamma}'\left(2{\beta}'+ {\gamma}'-{\alpha}'\right)-\frac{e^{2\alpha}}{2}\left(F-XF_X\right) + \frac{e^{2\alpha}}{2}\left[L(f)-\frac{q^2_{m}L_f}{e^{4\beta}}\right]=0, \\ \label{11}
-e^{2\alpha-2\beta} + {\beta}'' +{\beta}'\left(2{\beta}'+ {\gamma}'-{\alpha}'\right) -\frac{e^{2\alpha}}{2}\left[F-XF_X-L(f)\right]=0, \\ \label{12}
-e^{-2\beta} + e^{-2\alpha}{\beta}'\left({\beta}'+2{\gamma}'\right) -\frac{F}{2} + XF_X + \frac{L(f)}{2}=0.
\end{eqnarray}

The notation used here follows that used in reference \cite{KDJ}. The following coordinate transformation is defined: $u = x$, and the \textit{quasi-global} gauge $\alpha(u) + \gamma(u) = 0$ is employed. As a result, the line element in Eq. (\ref{6}) can be expressed in the following form:

\begin{eqnarray}\label{13}
ds^2= A\left(x\right)dt^2- \frac{dx^2}{A\left(x\right)} - \Sigma^2\left(x\right)d\Omega^2,
\end{eqnarray}
where the metric functions are defined as $A(x) = e^{2\gamma} = e^{-2\alpha}$ and $e^\beta = \Sigma(x)$. The equations of motion defined in Eqs. (\ref{9}-\ref{12}) can then be rewritten in the new coordinates. Combining Eqs. (\ref{10}-\ref{12}), we get the following expressions:

\begin{eqnarray}\label{14}
2A\frac{{\Sigma}''}{\Sigma} - XF_X =0, \\\label{15}
{A}''\Sigma^2 - A\left(\Sigma^2\right)''+ 2 -\frac{q^2_m{L_f}}{\Sigma^2} =0,
\end{eqnarray} where the primes now represent derivatives with respect to $x$.

The two remaining equations, Eq. (\ref{9}) and Eq. (\ref{12}), are rewritten in the new coordinates as
\begin{eqnarray}\label{16}
2\left(F_X{A\Sigma^2}\phi'\right)' - \Sigma^2F_\phi = 0, \\\label{17}
\frac{1}{\Sigma^2}\left(-1 + A'\Sigma'\Sigma + A{\Sigma'}^2\right) -\frac{F}{2} + XF_X + \frac{L(f)}{2} = 0.
\end{eqnarray}

It has been established in previous works \cite{CDJM1, CDJM2} that pursuing black-bounce solutions solely with the kinetic term of the k-essence function is not mathematically consistent. Therefore, when constructing these new solutions, we must incorporate a scalar potential given by $F(X) = F_0 X^n - 2V(\phi)$, where $F_0$ is a constant, $n$ is a real number and $V(\phi)$ is the potential function. Using Eq. (\ref{14}), we can derive a general expression for the scalar field that depends on both the angular metric function $\Sigma(x)$ and the metric function $A(x)$, unlike most studies, where the scalar field depends solely on the angular function and its derivatives \cite{KDJ, PRL, MM1}.

\section{First Model}\label{sec3}

In this section, we will consider the ingredients of the black-bounce solution proposed by Simpson-Visser in Ref. \cite{matt}. Specifically, we assume that the radius of the throat is exactly the magnetic charge $a=q_m$. Therefore, the metric functions are given by:

\begin{eqnarray}\label{18}
A(x)= 1- \frac{2m}{\sqrt{x^2+q^2_m}}, \qquad  \mbox{and} \qquad \Sigma(x)=\sqrt{x^2+q^2_m}.
\end{eqnarray}

In the following subsection, we will calculate all quantities present in the equations of motion (\ref{14}-\ref{17}) using metric functions Eq. (\ref{18}).

\subsection{case $n=\frac{1}{3}$}\label{sec31}

Considering the differential equation Eq. (\ref{14}) for the metric functions Eq. (\ref{18}), we can obtain the phantom scalar field for the configuration $n=\frac{1}{3}$ and the parameter $\eta=-1$. Therefore, its expression is given by

\begin{eqnarray}\label{19}
\phi(x)= \left(\frac{6q^2_m}{F_0}\right)^{3/2}\left[\frac{x}{4q^2_m{\Sigma}^4} + \frac{3x}{8q^4_m{\Sigma}^2} - \frac{2mx(15q^4_m+8x^4+20x^2{q^2_m})}{15q^6_m{\Sigma}^5} +\frac{3\arctan\left(\frac{x}{q_m}\right)}{8q^5_m}\right].
\end{eqnarray}
In the same way, using Eq. (\ref{15}), we can find the first electromagnetic quantity

\begin{eqnarray}\label{20}
L_f(x) = \frac{6m}{\sqrt{x^2+q^2_m}}.
\end{eqnarray}

Finally, using Eqs. (\ref{16})-(\ref{17}), we can find the scalar potential $V(x)$ as well as the electromagnetic quantity $L(f)$. Therefore, we have the following.

\begin{eqnarray}\label{21}
    V(x)&=& \frac{2q^2_m\left(5\Sigma-8m\right)}{5\Sigma^5},\\\label{22}
    L(x)&=& \frac{12mq^2_m}{5\Sigma^5}.
\end{eqnarray}

Electromagnetic quantities in general must obey the following relationship:

\begin{eqnarray}\label{23}
    \frac{\partial{L}}{\partial{x}}= \left(\frac{\partial{f}}{\partial{x}}\right)L_f.
\end{eqnarray} In this way, we can use the expression in Eq. (\ref{8}) for the Simpson-Visser area function and then write the electromagnetic quantity $L(x)$ as a function of the invariant $f=H_{\mu \nu}H^{\mu\nu}/4=\frac{q^2_m}{2\Sigma^4}$. Thus, we have:

\begin{eqnarray}\label{24}
    L(f)=\frac{24m{\sqrt[4]{2}}|f|^{5/4}}{5\sqrt{|q_m|}}.
\end{eqnarray}

To better visualize the form of the scalar potential in Eq. (\ref{21}), we can use an auxiliary variable defined by the transformation $\psi=\arctan\left(\frac{x}{q_m}\right)$,  where considering the asymptotic limits $x\rightarrow{\pm}\infty$ is equivalent to considering $\psi\rightarrow{\pm}\frac{\pi}{2}$. Thus, the potential in the new coordinates is given by:

\begin{eqnarray}\label{25}
    V(\psi)= \frac{2 \cos^4(\psi)}{q^2_m} - \frac{16m \cos^5(\psi)}{5q^3_m}.
\end{eqnarray}

It was verified in Ref. \cite{CDJM2} that regardless of the choice for the phantom scalar field configuration $n=1/3,1/5,1/7,\dots$, the energy conditions remain unchanged, preserving the results obtained in Ref. \cite{CDJM2}. The objective here is to examine in the following sections whether the energy conditions associated with the electromagnetic part are modified as we vary the power of the k-essence field.

The asymptotic form of the scalar field Eq. (\ref{19}) is written as:

\begin{eqnarray}\label{19a}
\phi\left(x\to{\infty}\right)=-\phi\left(x\to{-\infty}\right)= -\frac{\sqrt{3/2}\left(256m -45\pi{q_m}\right)}{20q^3_m},
\end{eqnarray} where $F_0=1$.

\begin{figure}[htb!]
\centering  
	\subfigure[]{\label{campoSV1}
	{\includegraphics[width=0.45\linewidth]{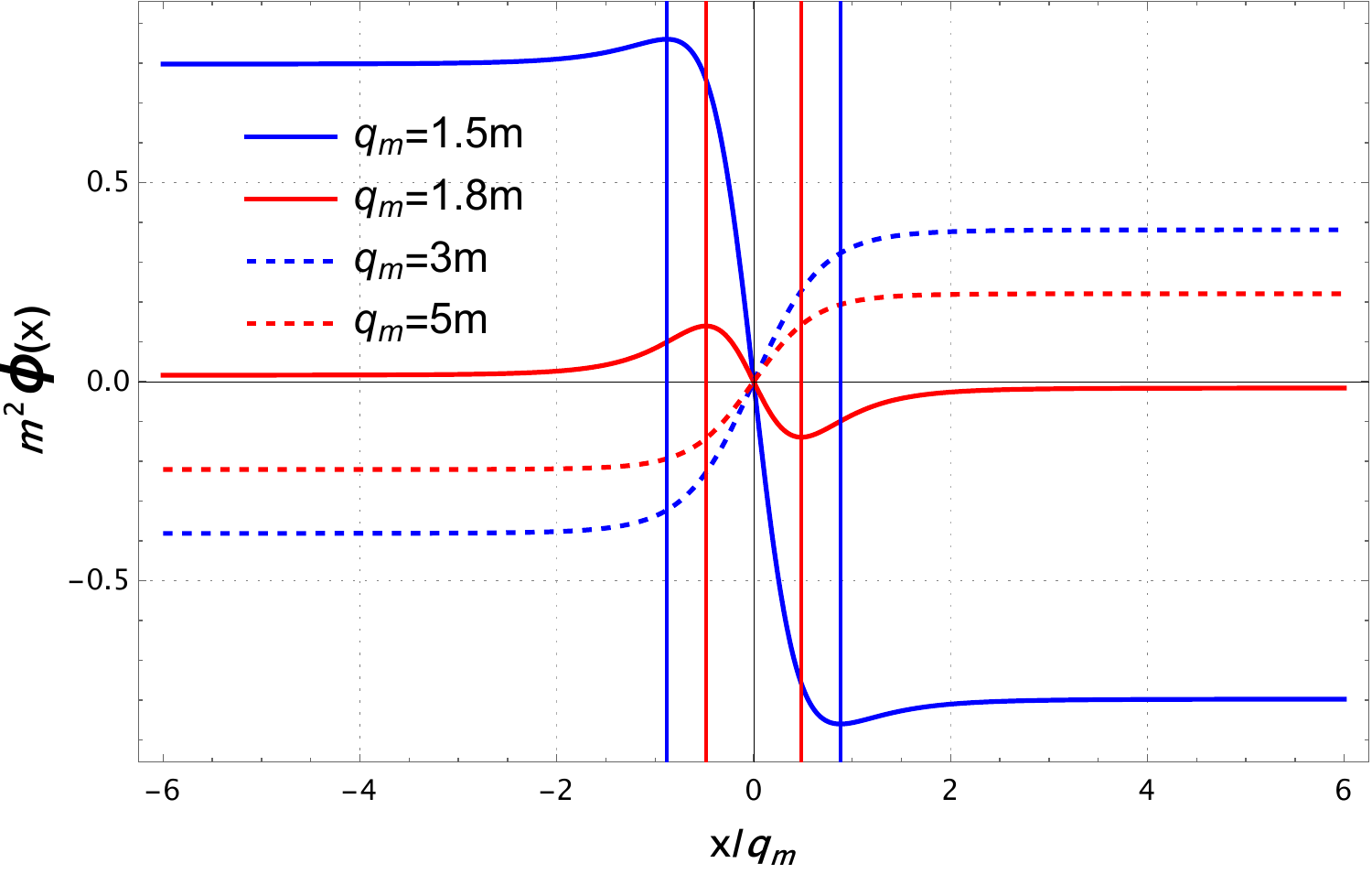}}}\qquad
	\subfigure[]{\label{poteSV1}
	{\includegraphics[width=0.45\linewidth]{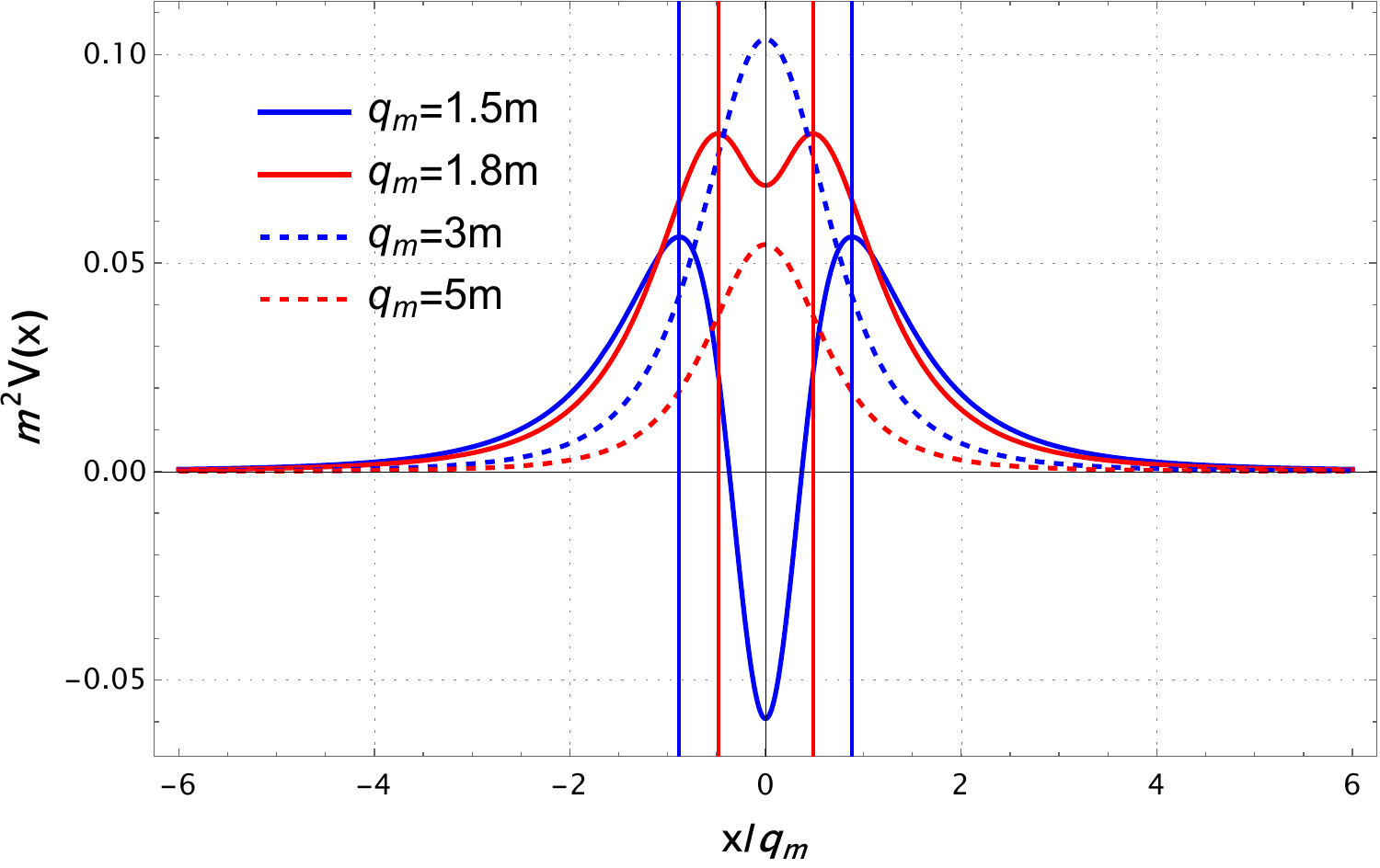}}} \qquad
	\subfigure[]{\label{Lf1SV}
	{\includegraphics[width=0.45\linewidth]{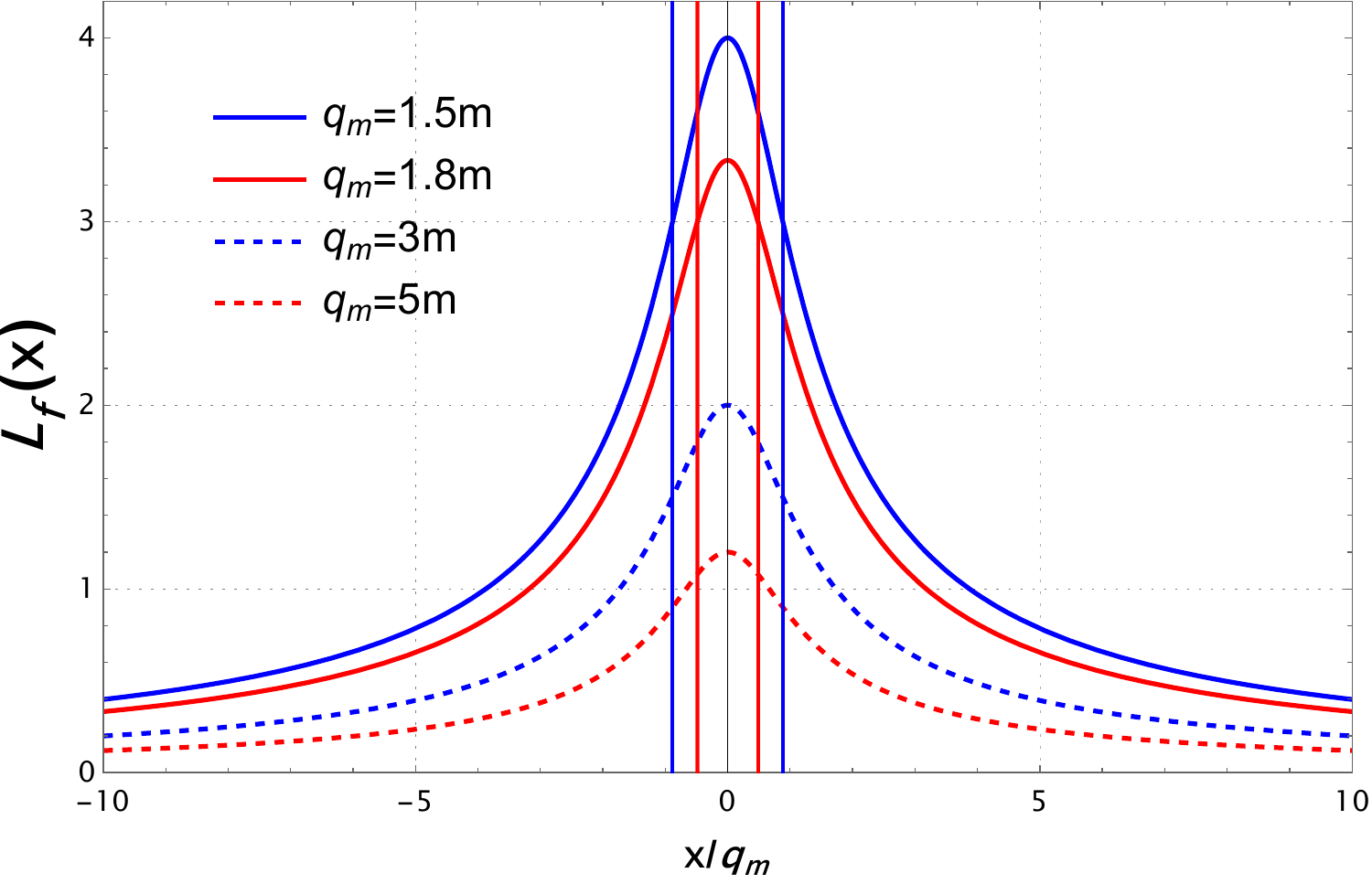}}} \qquad
	\subfigure[]{\label{L1SV}
	{\includegraphics[width=0.45\linewidth]{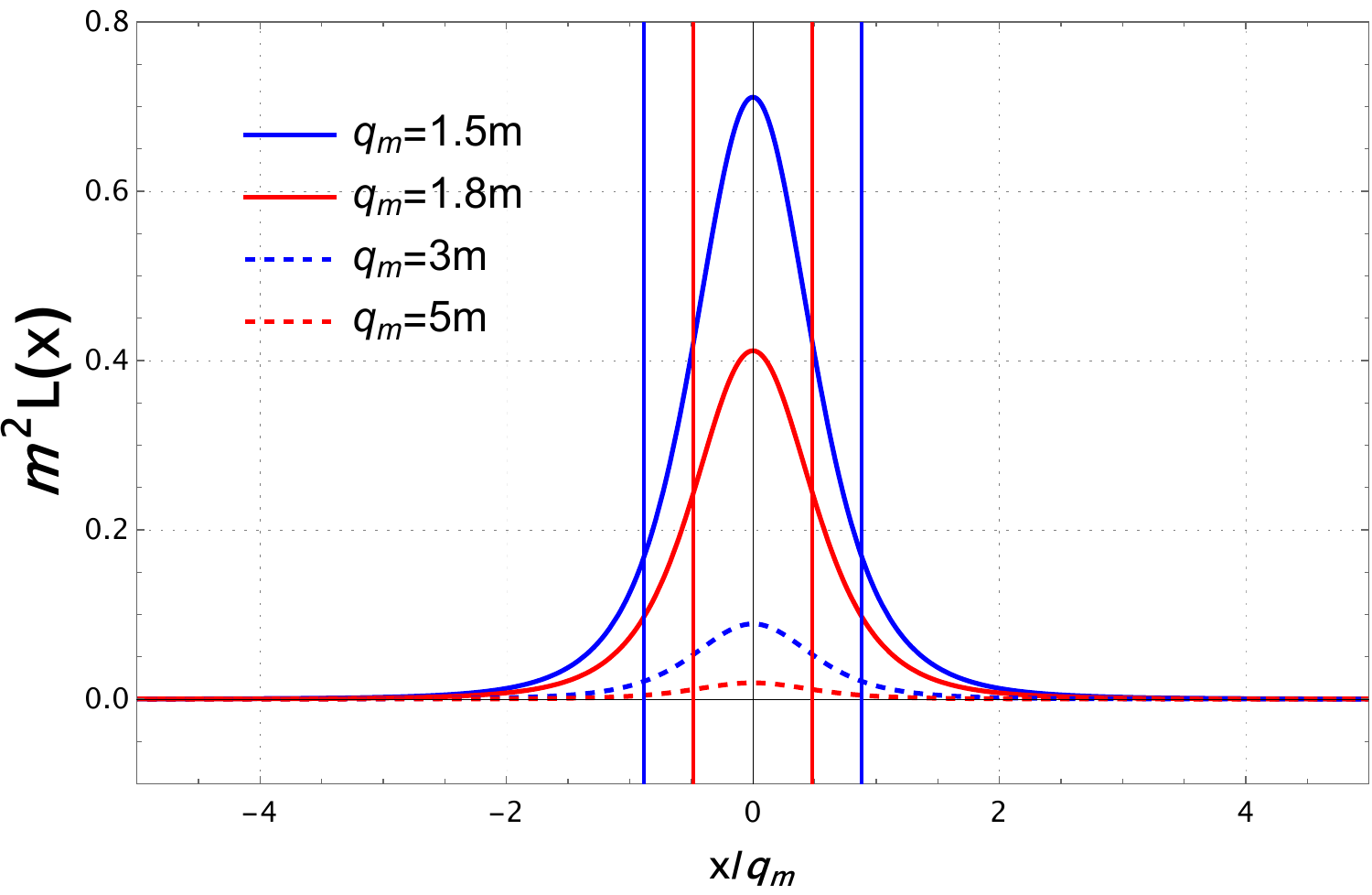}}}
\caption{The scalar field, given by Eq. (\ref{19}), is represented in a) for different charge values and considering regions inside and outside any possible horizon. Similarly, we have the potential, given by Eq. (\ref{20}), in b), and the electromagnetic functions, given by Eq. (\ref{21}) in c) and Eq. (\ref{22}) in d). All these are for $n=1/3$ setting, and we are adopting $F_0=1$. The vertical lines represent the positions of the event horizons. In the cases presented, horizons exist only for $q_m = 1.5m$ and $q_m = 1.8m$, because for $q_m > 2m$ there are no event horizons.}
\label{SV1}
\end{figure}

In Fig. \ref{SV1}, we have the graphical representation of the physical quantities obtained in the model above for the configuration of $n=1/3$. Regarding the scalar field (Fig. \ref{campoSV1}), looking at its asymptotic form Eq. (\ref{19a}), it is clear that there is a limit for charge values at which sign inversion occurs. For instance, for $x\to{+\infty}$, the field is positive for charge values around $q_m\approx{1.81m}$. Regarding the potential (Fig. \ref{poteSV1}), we can verify the same behavior presented in works \cite{CDJM1,CDJM2}, where, for certain charge values, already within the event horizon, the curves begin to create a minimum that grows as the charge becomes more internal. Likewise, we have the behavior of the electromagnetic functions in figures (Fig. \ref{Lf1SV}) and (Fig. \ref{L1SV}).

\subsection{case $n=\frac{1}{5}$}\label{sec32}

Considering the same procedure as in the previous section, we can explicitly obtain the same quantities, but now for a phantom scalar field configuration $n=1/5$. Therefore, we have:

 \begin{eqnarray}\label{26}
     \phi(x)&=&\left(\frac{10q^2_m}{F_0}\right)^{5/2}\left[\frac{35\arctan\left(\frac{x}{q_m}\right)}{128q^9_m} + \frac{93x}{128q^2_m{\Sigma^8}} +  \frac{35x^7}{128q^8_m{\Sigma^8}} +  \frac{385x^5}{384q^6_m{\Sigma^8}} +  \frac{511x^3}{384q^4_m{\Sigma^8}} +\frac{63m^2\arctan\left(\frac{x}{q_m}\right)}{64q^{11}_m}\right]  +  \nonumber   \\
   &+&  \left(\frac{10q^2_m}{F_0}\right)^{5/2}\left[ \frac{193m^2{x}}{64q^2_m\Sigma^{10}} + \frac{63m^2{x^9}}{64q^{10}_m\Sigma^{10}} + \frac{147m^2{x^7}}{32q^{8}_m\Sigma^{10}} + \frac{42m^2{x^5}}{5q^{6}_m\Sigma^{8}} + \frac{237m^2{x^3}}{32q^{4}_m\Sigma^{10}}\right] +  \nonumber   \\
    &-&  \left(\frac{10q^2_m}{F_0}\right)^{5/2}\left[\frac{4mx}{q^2_m\Sigma^9} + \frac{512mx^9}{315q^{10}_m\Sigma^9} +  \frac{256mx^7}{35q^{8}_m\Sigma^9} +  \frac{64mx^5}{5q^{6}_m\Sigma^9} +  \frac{32mx^3}{3q^{4}_m\Sigma^9}\right],
 \end{eqnarray}

\begin{eqnarray}\label{27}
    V(x)&=& \frac{4q^2_m}{\Sigma^4} -\frac{36mq^2_m}{5\Sigma^5}, \\\label{28}
    L(x)&=& \frac{12mq^2_m}{5\Sigma^5}.
\end{eqnarray}

An observation to be made is that the electromagnetic quantity $L_f$, obtained by Eq. (15), does not depend on the scalar field and is therefore the same as that obtained in the previous Section \ref{sec31}. Likewise, we can write the electromagnetic quantity $L(f)$ in terms of the invariant, as done in the previous section. Therefore, we have

\begin{eqnarray}\label{29}
    L(f)=\frac{24m\sqrt[4]{2}|f|^{5/4}}{5\sqrt{|q_m|}}.
\end{eqnarray}

We can also express the scalar potential for this field configuration using the most appropriate variables, as done in the previous section. Thus, we have
\begin{eqnarray}\label{30}
    V(\psi)= \frac{4\cos^4(\psi) }{q^2_m} -\frac{36m\cos^5(\psi)}{5q^3_m}.
\end{eqnarray}

The asymptotic form of the scalar field Eq. (\ref{26}) is written as:
\begin{eqnarray}\label{26a}
\phi\left(x\to{\infty}\right)=-\phi\left(x\to{-\infty}\right)= \frac{5 \sqrt{5/2} \left(39690 \pi  m^2-131072 m q_m+11025 \pi  q_m^2\right)}{2016 q_m^6},
\end{eqnarray} where $F_0=1$.

\begin{figure}[htb!]
\centering  
	\subfigure[]{\label{campoSV2}
	{\includegraphics[width=0.45\linewidth]{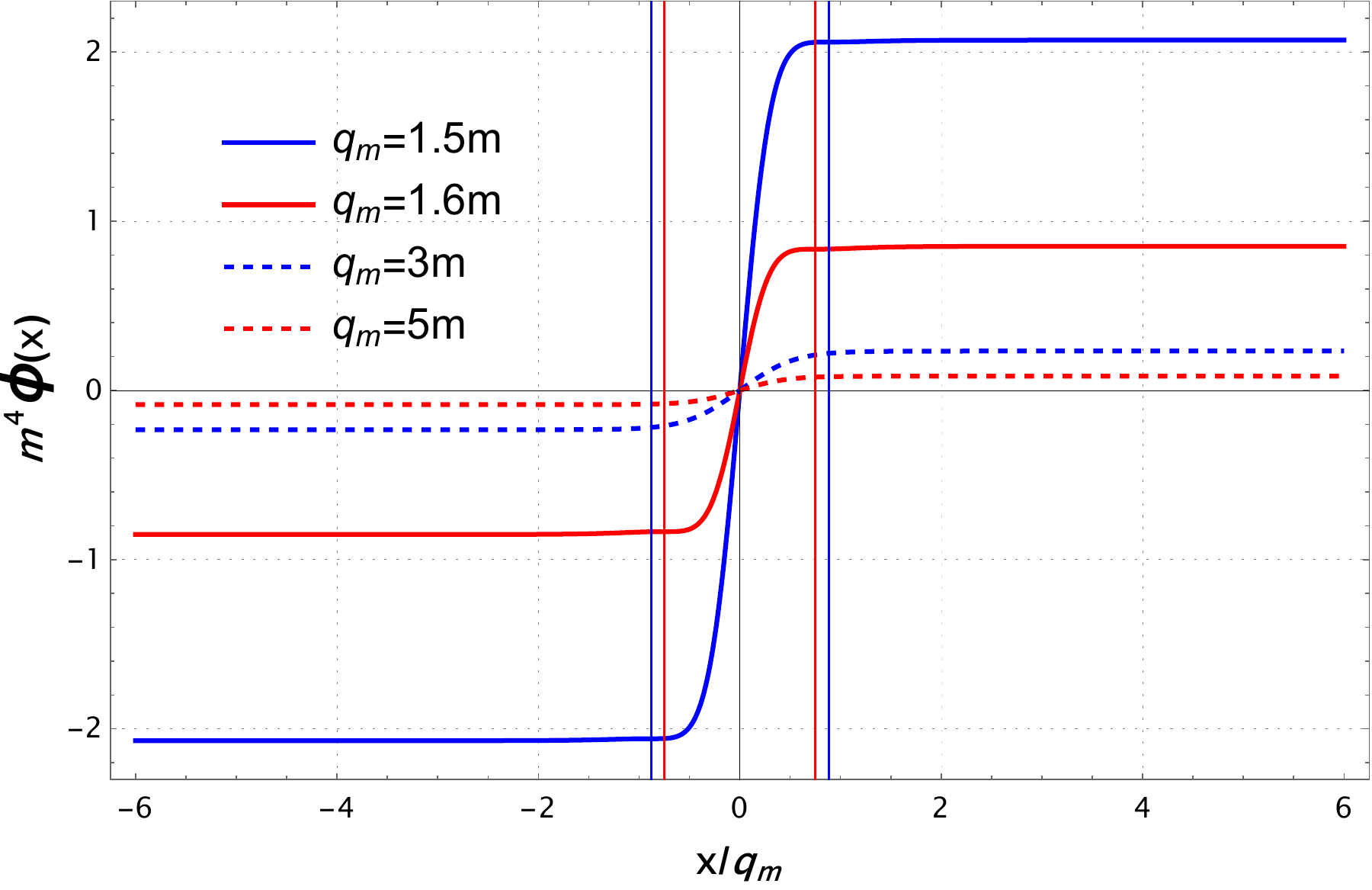}}}\qquad
	\subfigure[]{\label{poteSV2}
	{\includegraphics[width=0.45\linewidth]{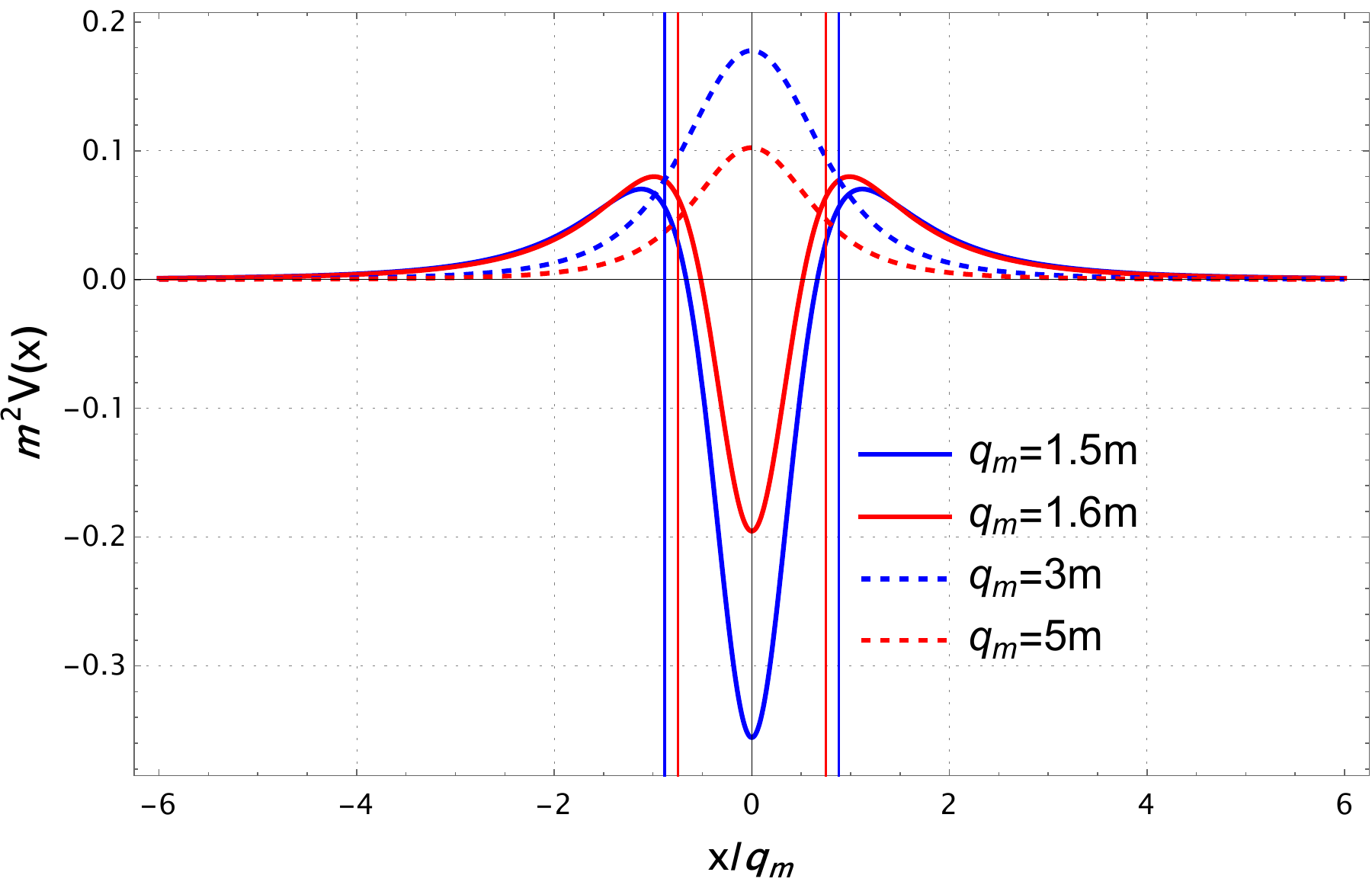}}}
\caption{Scalar field, Eq. (\ref{26}), in panel a) and potential, Eq. (\ref{27}), in panel b) for different charge values and considering regions inside and outside any possible horizon, for the configuration $n=1/5$. Here, we define $F_0=1$. The vertical lines represent the positions of the event horizons. In the cases presented, horizons exist only for $q_m = 1.5m$ and $q_m = 1.6m$, because for $q_m > 2m$ there are no event horizons.}
\label{SV2}
\end{figure}

In Fig. \ref{SV2}, we have the graphical representation of the scalar field, Eq. (\ref{26}), and potential, Eq. (\ref{27}), for the configuration $n=1/5$. As illustrated in Fig. \ref{campoSV2}, in the limit of $x\to\infty$ the scalar field grows for increasingly larger charge values and tends to zero for small values of the radial coordinate. The behavior of the scalar potential is qualitatively the same as the previous configuration (Section \ref{sec31}), changing only by a numerical shift (Fig. \ref{poteSV2}). From the asymptotic form for the scalar field in Eq. (\ref{26a}), we find that there are no real values for the charge that allow a sign change in this expression, contrary to what happens in the case $n = 1/3$.
 
\section{Second Model}\label{sec4}

In this section, we will work with the metric functions corresponding to a Bardeen-type spacetime as investigated in Refs. \cite{MM1,MM2}. In this model, as in the previous case, the wormhole throat coincides with the magnetic charge $q_m=a$. Therefore, we have:

\begin{eqnarray}\label{31}
    A(x)= 1-\frac{2mx^2}{(x^2+q^2_m)^{3/2}},\qquad \mbox{and} \qquad \Sigma(x)=\sqrt{x^2+q^2_m}.
\end{eqnarray}

As discussed in Refs. \cite{MM1,MM2}, this spacetime has four horizons, two event horizons and two Cauchy horizons. The extreme horizon is obtained for $q_m=q_{ext}=4m/3\sqrt{3}$ and the others are symmetric and labeled by $(-x_+,-x_C,x_C,x_+)$, where $x_{+}$ is the event horizon and $x_{C}$ is the Cauchy horizon.

\subsection{case $n=\frac{1}{3}$}\label{sec41}

Following the same steps carried out in Section \ref{sec31}, we can obtain all physical quantities of interest for this spacetime. Therefore, we have

\begin{eqnarray}\label{32}
\phi(x)&=&  \left(\frac{6q^2_m}{F_0}\right)^{3/2}\left[  \frac{5x}{8q^2_m{\Sigma^4}} +  \frac{3x^3}{8q^4_m{\Sigma^4}} - \frac{16mx^7}{105q^6_m{\Sigma^7}}  -\frac{2mx^3{q^4_m}}{3q^6_m{\Sigma^7}}  -\frac{8mx^5}{15q^4_m{\Sigma^7}} + \frac{3\arctan\left(\frac{x}{q_m}\right)}{8q^5_m} \right], \\\label{33}
L_f(x)&=& \frac{22mx^2}{{\Sigma^3}} - \frac{4mx^4}{q^2_m{\Sigma^3}} -\frac{4mq^2_m}{{\Sigma^3}} + \frac{4mx^2}{q^2_m{\Sigma}} , \\\label{34}
V(x)&=& \frac{2q^2_m}{{\Sigma^4}} - \frac{32mq^4_m}{35{\Sigma^7}} -\frac{16mx^2q^2_m}{5{\Sigma^7}}, \\\label{35}
L(x)&=&  \frac{64mq^4_m}{35{\Sigma^7}} +\frac{52mx^2q^2_m}{5{\Sigma^7}}.
\end{eqnarray}

Writing the electromagnetic quantity $L(x)$ in terms of the invariant $f$ and considering the appropriate change of variables for the scalar potential, we can rewrite them as:

\begin{eqnarray}\label{36}
    L(f)&=& \frac{52m (2|f|)^{5/4}}{5\sqrt{|q_m|}} - \frac{60m\sqrt{|q_m|} (2|f|)^{7/4}}{7} , \\\label{37}
    V(\psi)&=&  \frac{2\cos^4(\psi) }{q^2_m} - \frac{32m\cos^7(\psi) }{35q^3_m} - \frac{4m\cos^3(\psi)\sin^2(\psi) }{5q^3_m}.
\end{eqnarray}

The asymptotic form for the scalar field Eq. (\ref{32}) is written as:

\begin{eqnarray}\label{32ASSIM}
\phi\left(x\to\infty\right)= -\phi\left(x\to{-\infty}\right)=\frac{\sqrt{3/2}\left(315\pi{q_m}-256m\right)}{140q^3_m},
\end{eqnarray} where $F_0=1$.

\begin{figure}[htb!]
\centering  
	\subfigure[]{\label{campoBD1}
	{\includegraphics[width=0.45\linewidth]{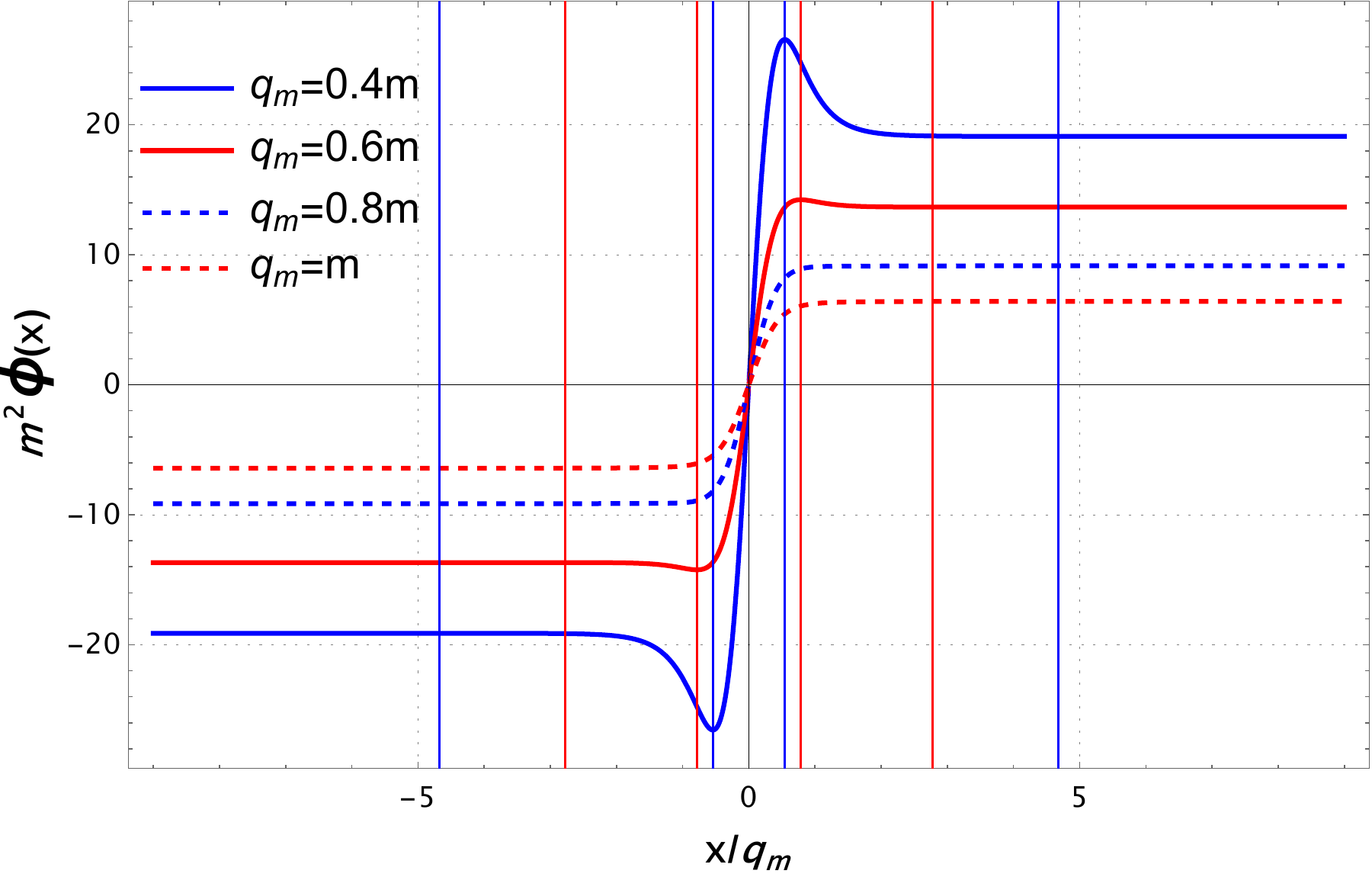}}}\qquad
	\subfigure[]{\label{poteBD1}
	{\includegraphics[width=0.45\linewidth]{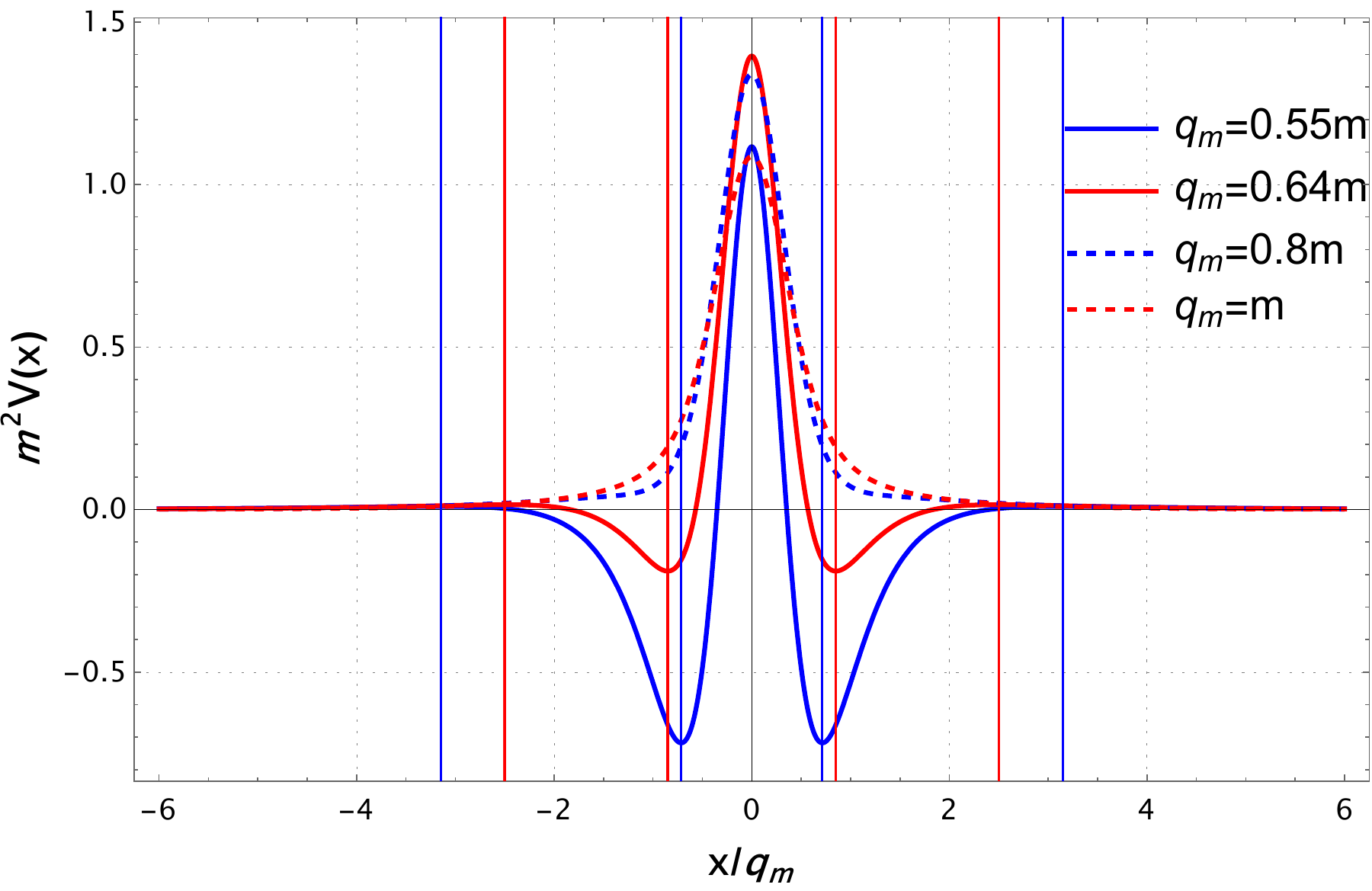}}} \qquad
	\subfigure[]{\label{Lf1BD}
	{\includegraphics[width=0.45\linewidth]{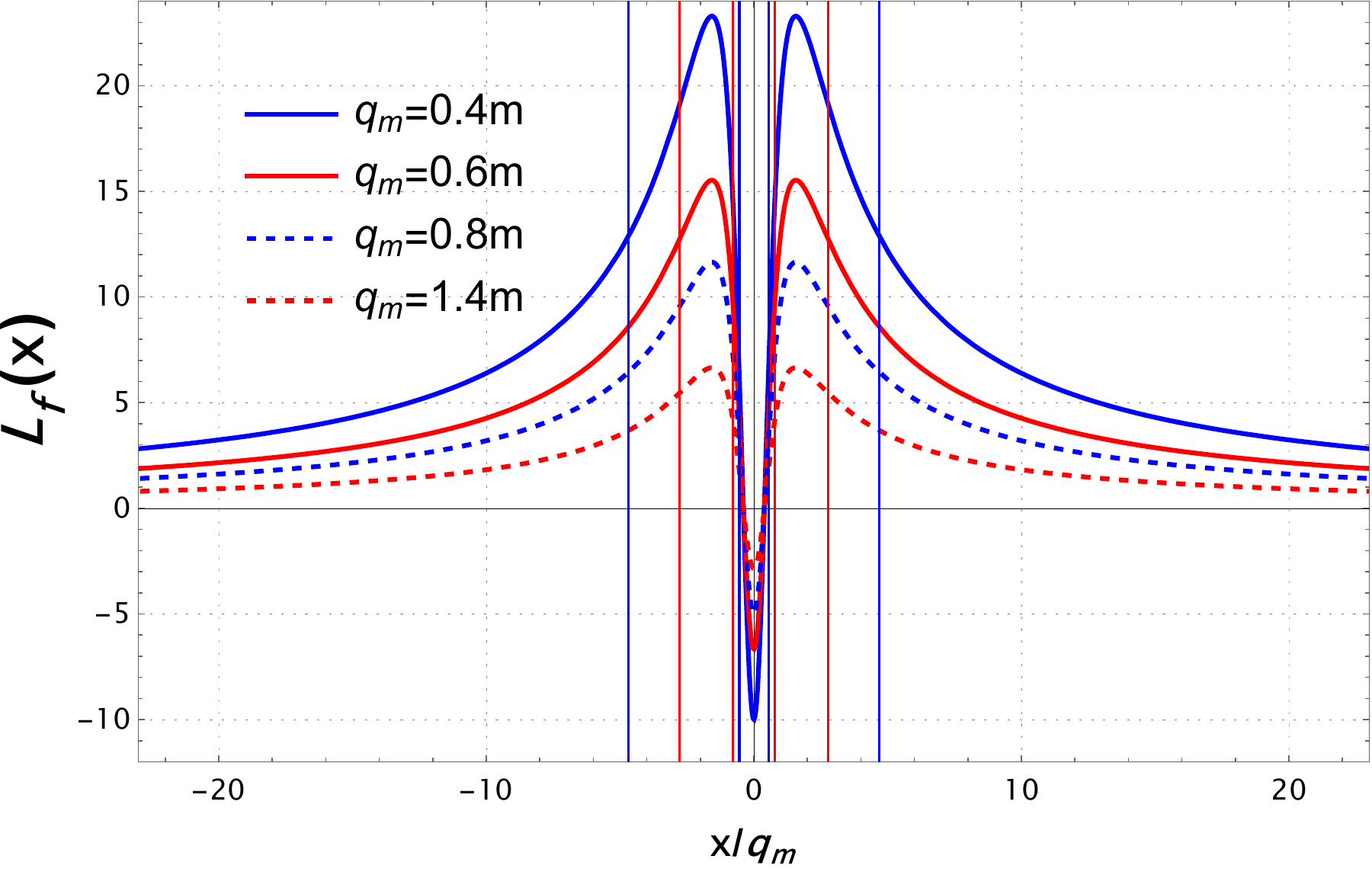}}} \qquad
	\subfigure[]{\label{L1BD}
	{\includegraphics[width=0.45\linewidth]{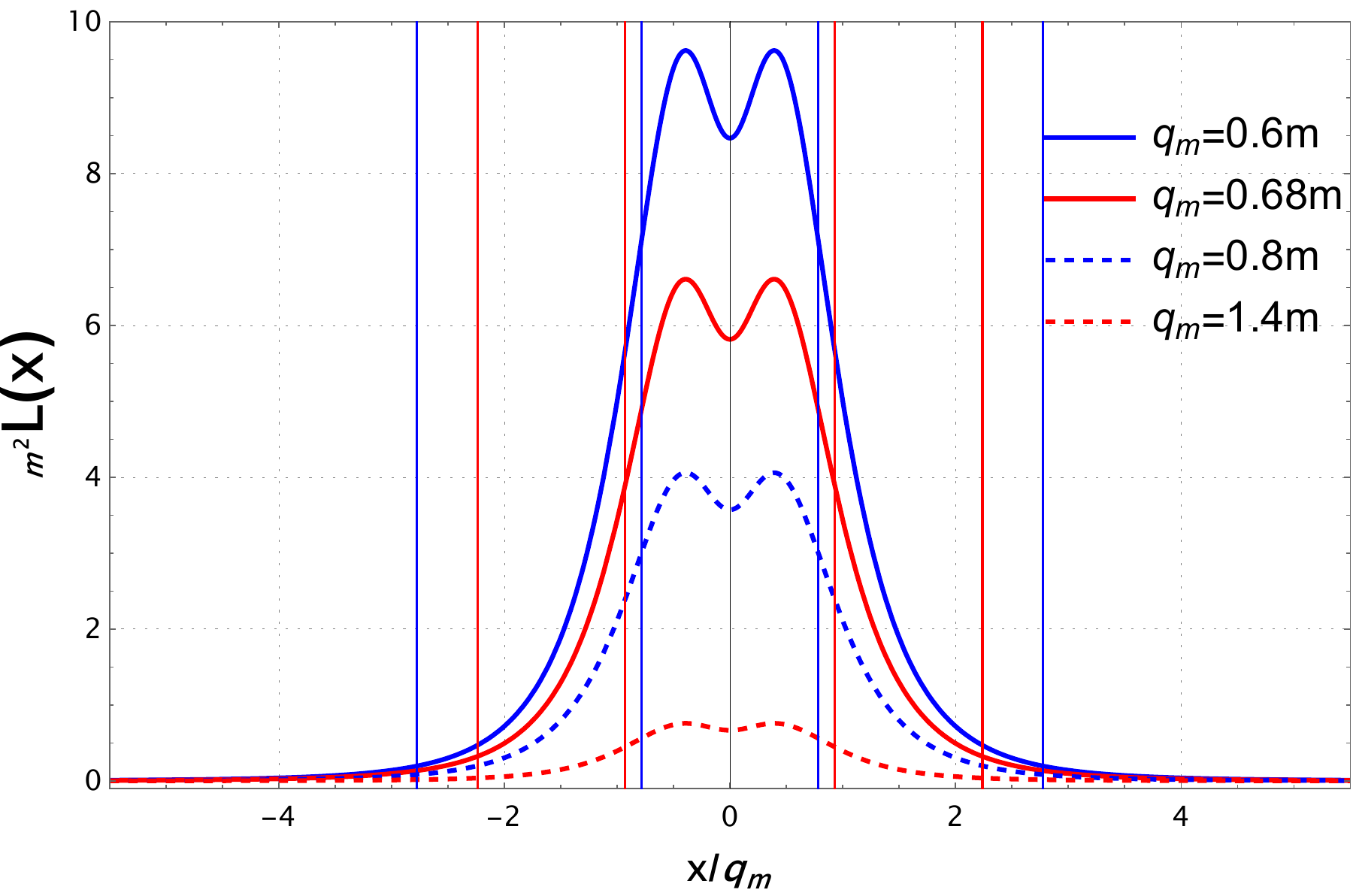}}}
\caption{Scalar field Eq. (\ref{32}) for some charge values and considering regions inside and outside any possible horizon is represented in a). Similarly, we have the potential Eq. (\ref{34}) b) and the electromagnetic functions Eq. (\ref{33}) c) and Eq. (\ref{35}) d). All for $n=1/3$ setting, adopting $F_0=1$. The vertical lines represent the positions of the horizons. For this model, there are four horizons: two event horizons (the outer horizons) and two Cauchy horizons (the inner horizons). In the cases presented, horizons exist only for $q_m < 4m/(3\sqrt{3})\approx0.7698m$.}
\label{BD1}
\end{figure}

In Fig. \ref{BD1}, we have a graphical representation of the physical quantities obtained above for the configuration $n=1/3$ for Bardeen-type spacetime. In Fig. \ref{campoBD1}, we show the behavior of the scalar field, Eq. (\ref{32}), in all regions of the spacetime, considering different charge values. For some values of charge, $q_m>q_{ext}=\frac{4m}{3\sqrt{3}}$, there is no event horizon. Note in the asymptotic form of the scalar field Eq. (\ref{32ASSIM}) that for a certain limiting value of charge within the horizon, around $q_m\approx{0.26m}$, the scalar field tends to reverse sign. 

For the scalar potential, Fig. \ref{poteBD1}, we observe a behavior similar to the cases previously investigated. For regions outside the external horizon, the potential behaves like a barrier. For regions within the horizon, the potential starts to exhibit two symmetrical minima that tend to grow as the charge becomes more internal to extreme radius. The electromagnetic quantities Eq. (\ref{33}) and Eq. (\ref{35}) exhibit behavior similar to that presented in Section \ref{sec31} and can be visualized in Figs. \ref{Lf1BD} and \ref{L1BD}.

\subsection{case $n=\frac{1}{5}$}\label{sec42}

Following the same steps as in the previous Section \ref{sec41}, we can obtain the same quantities, but now for the phantom field power $n=1/5$. In this way, we have:

\begin{eqnarray}\label{38}
\phi(x)&=& \left(\frac{10q^2_m}{F_0}\right)^{5/2}\left[\frac{35\arctan\left(\frac{x}{q_m}\right)}{128q^9_m} + \frac{93x}{128q^2_m{\Sigma^8}} +  \frac{35x^7}{128q^8_m{\Sigma^8}} +  \frac{385x^5}{384q^6_m{\Sigma^8}} +  \frac{511x^3}{384q^4_m{\Sigma^8}} + \frac{9m^2\arctan\left(\frac{x}{q_m}\right)}{512q^{11}_m} \right]  +  \nonumber   \\
&+& \left(\frac{10q^2_m}{F_0}\right)^{5/2}\left[-\frac{9m^2q^2_m{x}}{512\Sigma^{14}} + \frac{1199m^2x^5}{2560{q^2_m}\Sigma^{14}} - \frac{15m^2x^3}{128\Sigma^{14}} + \frac{9m^2x^{13}}{512{q^{10}_m}\Sigma^{14}} + \frac{15m^2x^{11}}{128{q^8_m}\Sigma^{14}} + \frac{849m^2x^9}{2560{q^6_m}\Sigma^{14}} + \frac{18m^2x^7}{35{q^4_m}\Sigma^{14}}\right] +  \nonumber   \\
&-&\left(\frac{10q^2_m}{F_0}\right)^{5/2}\left[\frac{4m{x^3}}{3q^2_m\Sigma^{11}} + \frac{512m{x^{11}}}{3465q^{10}_m\Sigma^{11}} + \frac{256m{x^9}}{315q^8_m\Sigma^{11}} + \frac{64m{x^7}}{35q^6_m\Sigma^{11}} + \frac{32m{x^5}}{15q^4_m\Sigma^{11}}\right],
\end{eqnarray}

\begin{eqnarray}\label{39}
        V(x)&=& \frac{8mq^2_m}{5\Sigma^5} +  \frac{4q^2_m}{\Sigma^4} -  \frac{88mq^4_m}{35\Sigma^7} -  \frac{44mx^2q^2_m}{5\Sigma^7}, \\\label{40}
        L(x)&=& \frac{68mx^2q^2_m}{5\Sigma^7} + \frac{176mq^4_m}{35\Sigma^7} -\frac{16mq^2_m}{5\Sigma^5}.
\end{eqnarray}

Finally, by expressing the potential Eq. (\ref{39}) in the new coordinates, as well as the electromagnetic quantity Eq. (\ref{40}) in terms of the invariant $f$, we have:

\begin{eqnarray}\label{41}
    V(\psi)&=& \frac{8m\cos^5(\psi) }{5q^3_m} +  \frac{4\cos^4(\psi) }{q^2_m} - \frac{88m\cos^7(\psi) }{35q^3_m} -\frac{44m\cos^5(\psi)\sin^2(\psi) }{5q^3_m}, \\\label{42}
    L(f)&=& \frac{52m(2|f|)^{5/4}}{5\sqrt{|q_m|}} - \frac{60m\sqrt{|q_m|}(2|f|)^{7/4}}{7} .
\end{eqnarray}

The asymptotic form for the scalar field Eq. (\ref{38}) is written as:
\begin{eqnarray}\label{38ASSIM}
\phi\left(x\to\infty\right)= -\phi\left(x\to{-\infty}\right)=\frac{5 \sqrt{5/2} \left(31185 \pi  m^2-524288 m q_m+485100 \pi  q_m^2\right)}{88704 q_m^6},
\end{eqnarray} where $F_0=1$.

\begin{figure}[htb!]
\centering  
	\subfigure[]{\label{campoBD2}
	{\includegraphics[width=0.45\linewidth]{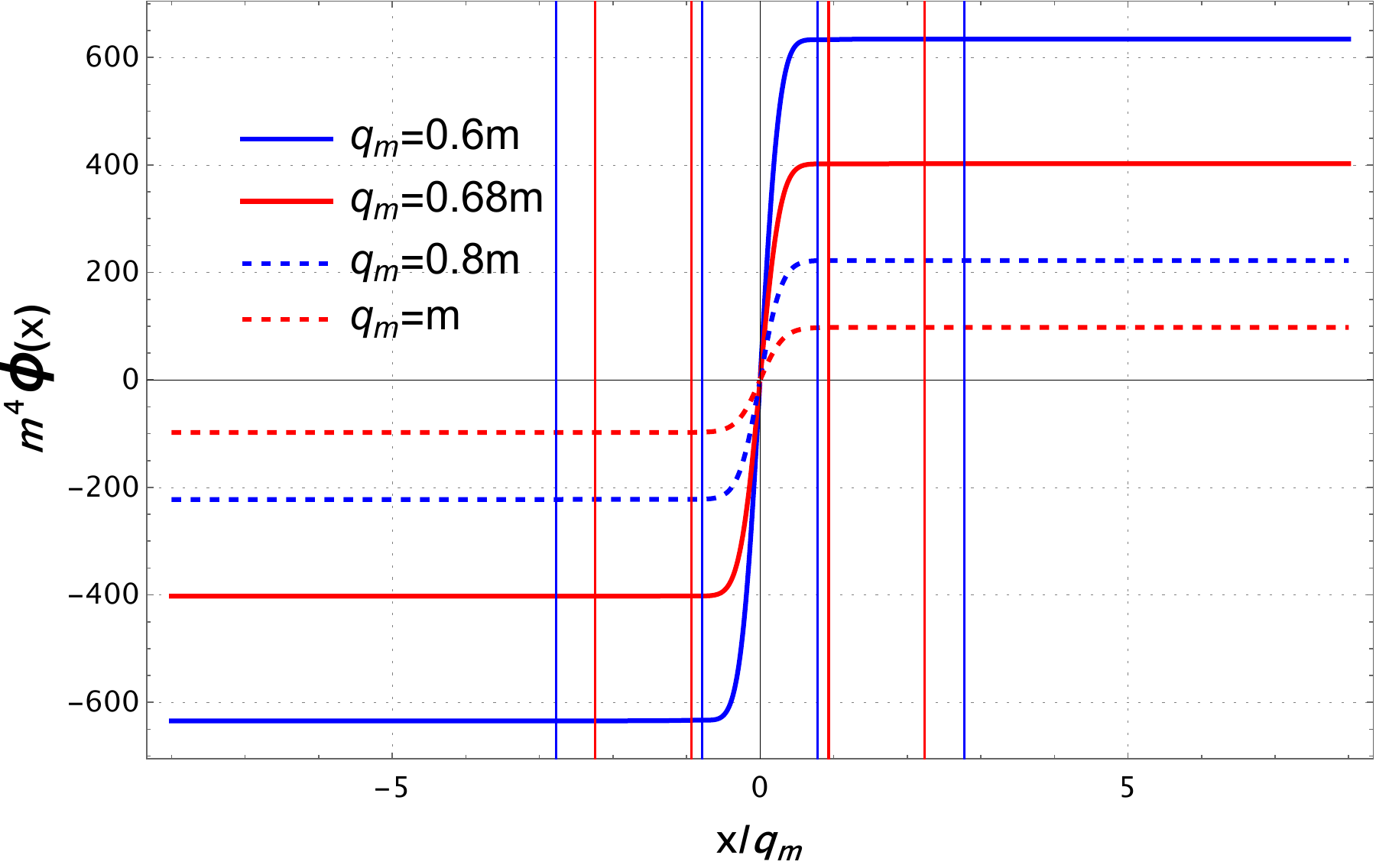}}}\qquad
	\subfigure[]{\label{poteBD2}
	{\includegraphics[width=0.45\linewidth]{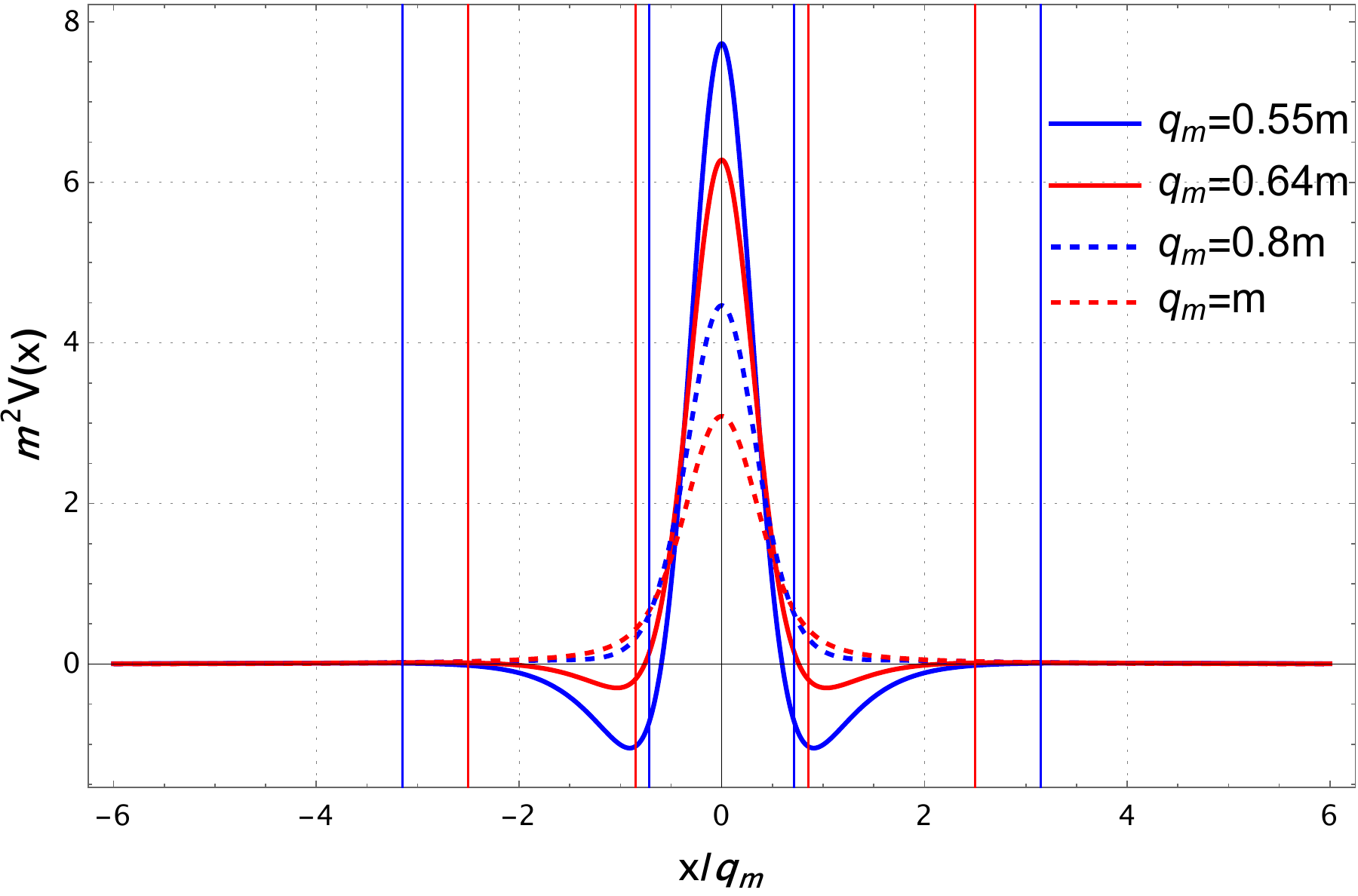}}}
\caption{Scalar field, Eq. (\ref{38}), in panel a) and potential, Eq. (\ref{39}), in panel b) with different charge values and considering regions inside and outside any possible horizon, for the configuration $n=1/5$ . Here, we define $F_0=1$. The vertical lines represent the positions of the horizons. For this model, there are four horizons: two event horizons (the outer horizons) and two Cauchy horizons (the inner horizons). In the cases presented, horizons exist only for $q_m < 4m/(3\sqrt{3})\approx0.7698m$.}
\label{BD2}
\end{figure}

In Fig. \ref{BD2}, we have the graphical representation for the scalar field Eq. (\ref{38}) and potential Eq. (\ref{39}) with a configuration $n=1/5$. Qualitatively, the scalar field Fig. \ref{campoBD2} and the potential Fig. \ref{poteBD2} exhibit similar behavior to the previously investigated cases, with only minor numerical shifts. From the asymptotic form for the scalar field in Eq. (\ref{38ASSIM}), we find that there are no real values for the charge that allow a sign change in this expression, identical to what was observed in the Simpson-Visser case with $n = 1/5$.

An important observation is that the physical quantities associated with electromagnetism are the same regardless of the power of the k-essence field. Surprisingly, these quantities are the same as those obtained for the canonical case \cite{MM1}. Logically, these electromagnetic functions, which remain unchanged about the k-essence field, are specific to each model of interest. As is the case with the functions Eqs. (\ref{24}) and (\ref{29}) which refer to the Simpson-Visser model and are the same expressions obtained for the canonical scalar field in \cite{MM1}. The other electromagnetic quantity Eq. (\ref{20}) is also the same for any power of the k-essence field and coincides with the canonical case. This fact occurs because it does not depend on the scalar field and is obtained directly through Eq. (\ref{15}).

The same analysis can be extended to the second model (Bardeen-type) when comparing the functions Eqs. (\ref{36}) and (\ref{42}) as well as the expression Eq. (\ref{33}) which are the same found for the canonical scalar field \cite{MM1}.

The fact that the electromagnetic quantities remain unchanged is compensated by changes in the phantom scalar field and scalar potential, as a kind of counterterms.





\section{General Energy Conditions}\label{sec5}

We know that black-bounce solutions violate the null energy condition, which must violate all other energy conditions \cite{matt,LZ}. In this section, we will make the generic construction for the energy conditions, that is, just considering the static and spherically symmetric metric Eq. (\ref{13}), as well as the stress-energy tensor defined in Eqs. (\ref{4}-\ref{5}). The energy conditions are defined by the combination of energy density and radial and tangential pressures as\footnote{In the notation we follow, NEC is used for the null energy condition, WEC for the weak energy condition, SEC for the strong energy condition, and DEC for the dominant energy condition. The subscript $1$ is used when we have the energy density with radial pressure, $2$ when we have the energy density with tangential pressure and $3$ when we have both pressures or just the energy density.}

\begin{eqnarray}\label{43}
NEC^{\phi,EM}_{1,2}&=&WEC^{\phi,EM}_{1,2}=SEC^{\phi,EM}_{1,2} \Longleftrightarrow \rho^{\phi,EM} + p^{\phi,EM}_{1,2} \geq 0, \\\label{44}
SEC^{\phi,EM}_3 &\Longleftrightarrow & \rho^{\phi,EM} + p^{\phi,EM}_{1} + 2p^{\phi,EM}_{2} \geq 0, \\\label{45}
DEC^{\phi,EM}_{1,2} &\Longleftrightarrow &  \rho^{\phi,EM} + p^{\phi,EM}_{1,2} \geq 0  \qquad    \mbox{and} \qquad \rho^{\phi,EM} - p^{\phi,EM}_{1,2} \geq 0 , \\\label{46}
DEC^{\phi,EM}_3&=&WEC^{\phi,EM}_{3} \Longleftrightarrow   \rho^{\phi,EM}  \geq 0 ,
\end{eqnarray} with the subindices $\phi$ and $EM$ in the energy conditions above, representing the scalar field and the electromagnetic field.

The stress-energy tensor is written as
\begin{eqnarray}\label{47}
\tensor{T}{^\mu}_{\nu}= {\rm diag}\left[\rho^{\phi,EM},-p^{\phi,EM}_1,-p^{\phi,EM}_2,-p^{\phi,EM}_2\right],
\end{eqnarray} where the energy density is denoted by $\rho^{\phi,EM}$, while $p^{\phi,EM}_1$ and $p^{\phi,EM}_2$ represent the radial and tangential pressures, respectively. This expression for the stress-energy tensor above, Eq. (\ref{47}), refers to the part outside the event horizon, $A>0$. Within the event horizon, $A<0$, we have:

\begin{eqnarray}\label{48}
\tensor{T}{^\mu}_{\nu}={\rm diag}\left[-p^{\phi,EM}_1,\rho^{\phi,EM},-p^{\phi,EM}_2,-p^{\phi,EM}_2\right].
\end{eqnarray}

\subsection{Energy conditions for $n=1/3$}\label{sec51}

As the energy conditions depend directly on the scalar field, we will consider these components of the stress-energy tensor to the field configuration where $n=1/3$. For the region $A>0$, the components are defined as:

\begin{eqnarray}\label{49}
    \rho^\phi &=& -\frac{3A{\Sigma}''}{\Sigma} + V(x), \quad \quad \rho^{EM}= \frac{L(x)}{2}, \\\label{50}
    {p}^\phi_1 &=& \frac{A{\Sigma}''}{\Sigma} - V(x), \quad \quad  {p}^{EM}_1= -\frac{L(x)}{2}, \\\label{51}
    {p}^\phi_2 &=&\frac{3A{\Sigma}''}{\Sigma} - V(x), \quad \quad  {p}^{EM}_2= -\frac{L(x)}{2} + \frac{q^2_m{L_f(x)}}{2\Sigma^4}.
\end{eqnarray}

The components of the stress-energy tensor for the region $A<0$ are defined by:

\begin{eqnarray}\label{52}
    \rho^\phi &=& -\frac{A{\Sigma}''}{\Sigma} + V(x), \quad \quad \rho^{EM}= \frac{L(x)}{2}, \\\label{53}
    {p}^\phi_1 &=& \frac{3A{\Sigma}''}{\Sigma} - V(x), \quad \quad  {p}^{EM}_1= -\frac{L(x)}{2}, \\\label{54}
    {p}^\phi_2 &=&  \frac{3A{\Sigma}''}{\Sigma} - V(x), \quad \quad  {p}^{EM}_2= -\frac{L(x)}{2} + \frac{q^2_m{L_f(x)}}{2\Sigma^4}.
\end{eqnarray}

The energy conditions for the region outside the event horizon $A>0$ are:

\begin{eqnarray}\label{55}
NEC^{\phi}_{1}&=&WEC^{\phi}_{1}=SEC^{\phi}_{1}= -\frac{2A{\Sigma}''}{\Sigma} \geq 0, \\\label{56}
NEC^{\phi}_{2}&=&WEC^{\phi}_{2}=SEC^{\phi}_{2}= 0,  \\\label{57}
SEC^{\phi}_{3}&=& \frac{4A{\Sigma}''}{\Sigma} -2V(x) \geq 0, \\\label{58}
DEC^{\phi}_{1}&=&  - \frac{4A{\Sigma}''}{\Sigma} + 2V(x) \geq 0,  \\\label{59}
DEC^{\phi}_{2}&=&  - \frac{6A{\Sigma}''}{\Sigma} + 2V(x) \geq 0,  \\\label{60}
DEC^{\phi}_{3}&=& WEC^{\phi}_{3}=  - \frac{3A{\Sigma}''}{\Sigma} + V(x) \geq 0.
\end{eqnarray}

Likewise, for the electromagnetic part

\begin{eqnarray}\label{61}
NEC^{EM}_{1}&=&WEC^{EM}_{1}=SEC^{EM}_{1}= 0, \\\label{62}
NEC^{EM}_{2}&=&WEC^{EM}_{2}=SEC^{EM}_{2}= \frac{q^2_m{L_f(x)}}{2\Sigma^4} \geq 0,  \\\label{63}
SEC^{EM}_{3}&=& - L(x) + \frac{q^2_m{L_f(x)}}{\Sigma^4} \geq 0, \\\label{64}
DEC^{EM}_{1}&=& L(x) \geq 0,  \\\label{65}
DEC^{EM}_{2}&=&  L(x) - \frac{q^2_m{L_f(x)}}{2\Sigma^4}\geq 0,  \\\label{66}
DEC^{EM}_{3}&=& WEC^{EM}_{3}=   \frac{L(x)}{2} \geq 0.
\end{eqnarray}

Performing the same analysis now within the horizon $A<0$, we have:

\begin{eqnarray}\label{67}
NEC^{\phi}_{1}&=&WEC^{\phi}_{1}=SEC^{\phi}_{1}= \frac{2A{\Sigma}''}{\Sigma} \geq 0, \\\label{68}
NEC^{\phi}_{2}&=&WEC^{\phi}_{2}=SEC^{\phi}_{2}=  \frac{2A{\Sigma}''}{\Sigma} \geq 0,  \\\label{69}
SEC^{\phi}_{3}&=& \frac{8A{\Sigma}''}{\Sigma} -2V(x) \geq 0, \\\label{70}
DEC^{\phi}_{1}&=&  - \frac{4A{\Sigma}''}{\Sigma} + 2V(x) \geq 0,  \\\label{71}
DEC^{\phi}_{2}&=&  - \frac{4A{\Sigma}''}{\Sigma} + 2V(x) \geq 0,  \\\label{72}
DEC^{\phi}_{3}&=& WEC^{\phi}_{3}=  - \frac{A{\Sigma}''}{\Sigma} + V(x) \geq 0.
\end{eqnarray}

The electromagnetic part within the horizon is the same expressions obtained above in Eqs. (\ref{61}-\ref{66}).


\subsection{Energy conditions for $n=1/5$}\label{sec52}

The components of the stress-energy tensor for the region outside the event horizon $A>0$ for the scalar field configuration in question are defined as

\begin{eqnarray}\label{73}
    \rho^\phi &=& -\frac{5A{\Sigma}''}{\Sigma} + V(x), \quad \quad \rho^{EM}= \frac{L(x)}{2}, \\\label{74}
    {p}^\phi_1 &=& \frac{3A{\Sigma}''}{\Sigma} - V(x), \quad \quad  {p}^{EM}_1= -\frac{L(x)}{2}, \\\label{75}
    {p}^\phi_2 &=& \frac{5A{\Sigma}''}{\Sigma} - V(x), \quad \quad  {p}^{EM}_2= -\frac{L(x)}{2} + \frac{q^2_m{L_f(x)}}{2\Sigma^4}.
\end{eqnarray}

Likewise for the internal region $A<0$

\begin{eqnarray}\label{76}
    \rho^\phi &=& -\frac{3A{\Sigma}''}{\Sigma} + V(x), \quad \quad \rho^{EM}= \frac{L(x)}{2}, \\\label{77}
    {p}^\phi_1 &=& \frac{5A{\Sigma}''}{\Sigma} - V(x), \quad \quad  {p}^{EM}_1= -\frac{L(x)}{2}, \\\label{78}
    {p}^\phi_2 &=& \frac{5A{\Sigma}''}{\Sigma} - V(x), \quad \quad  {p}^{EM}_2= -\frac{L(x)}{2} + \frac{q^2_m{L_f(x)}}{2\Sigma^4}.
\end{eqnarray}

The energy conditions external to the horizon $A>0$ for this field configuration are given by

\begin{eqnarray}\label{79}
NEC^{\phi}_{1}&=&WEC^{\phi}_{1}=SEC^{\phi}_{1}= -\frac{2A{\Sigma}''}{\Sigma} \geq 0, \\\label{80}
NEC^{\phi}_{2}&=&WEC^{\phi}_{2}=SEC^{\phi}_{2}= 0,  \\\label{81}
SEC^{\phi}_{3}&=& \frac{8A{\Sigma}''}{\Sigma} -2V(x) \geq 0, \\\label{82}
DEC^{\phi}_{1}&=&  - \frac{8A{\Sigma}''}{\Sigma} + 2V(x) \geq 0,  \\\label{83}
DEC^{\phi}_{2}&=&  - \frac{10A{\Sigma}''}{\Sigma} + 2V(x) \geq 0,  \\\label{84}
DEC^{\phi}_{3}&=& WEC^{\phi}_{3}=  - \frac{5A{\Sigma}''}{\Sigma} + V(x) \geq 0.
\end{eqnarray}

Likewise, for the electromagnetic part

\begin{eqnarray}\label{85}
NEC^{EM}_{1}&=&WEC^{EM}_{1}=SEC^{EM}_{1}= 0, \\\label{86}
NEC^{EM}_{2}&=&WEC^{EM}_{2}=SEC^{EM}_{2}= \frac{q^2_m{L_f(x)}}{2\Sigma^4} \geq 0,  \\\label{87}
SEC^{EM}_{3}&=& - L(x) + \frac{q^2_m{L_f(x)}}{\Sigma^4} \geq 0, \\\label{88}
DEC^{EM}_{1}&=& L(x) \geq 0,  \\\label{89}
DEC^{EM}_{2}&=&  L(x) - \frac{q^2_m{L_f(x)}}{2\Sigma^4}\geq 0,  \\\label{90}
DEC^{EM}_{3}&=& WEC^{EM}_{3}=   \frac{L(x)}{2} \geq 0.
\end{eqnarray}

The energy conditions where $A<0$ for this field configuration are given by

\begin{eqnarray}\label{91}
NEC^{\phi}_{1}&=&WEC^{\phi}_{1}=SEC^{\phi}_{1}= \frac{2A{\Sigma}''}{\Sigma} \geq 0, \\\label{92}
NEC^{\phi}_{2}&=&WEC^{\phi}_{2}=SEC^{\phi}_{2}= \frac{2A{\Sigma}''}{\Sigma} \geq 0,  \\\label{93}
SEC^{\phi}_{3}&=& \frac{12A{\Sigma}''}{\Sigma} -2V(x) \geq 0, \\\label{94}
DEC^{\phi}_{1}&=&  - \frac{8A{\Sigma}''}{\Sigma} + 2V(x) \geq 0,  \\\label{95}
DEC^{\phi}_{2}&=&  - \frac{8A{\Sigma}''}{\Sigma} + 2V(x) \geq 0,  \\\label{96}
DEC^{\phi}_{3}&=& WEC^{\phi}_{3}=  - \frac{3A{\Sigma}''}{\Sigma} + V(x) \geq 0.
\end{eqnarray}

The electromagnetic parts within the horizon are the same expressions as obtained above in Eqs. (\ref{85}-\ref{90}).

Analyzing the energy conditions in a generic way, we can emphasize that the relations obtained above in relation to the phantom scalar field remain unchanged for higher powers of the k-essence field $n=1/5,1/7,1/9, \dots $, and it is sufficient to analyze the lowest power case $n=1/3$ for the specific model of interest. This result corroborates the analyses carried out in \cite{CDJM2}.

\section{Analysis of energy conditions for each specific model}\label{sec6}

\subsection{Simpson-Visser model}\label{sec61}

The energy conditions for the scalar field and electromagnetism, in the region external to the event horizon, $A>0$, and considering the $n=1/3$ configuration, are set out explicitly below, where the quantities Eqs. (\ref{19}-\ref{22}) will be used.

\begin{eqnarray}\label{97}
NEC^{\phi}_{1}&=& -\frac{2 q^2_m \left(\sqrt{q^2_m+x^2}-2 m\right)}{\left(q^2_m+x^2\right)^{5/2}}  \geq 0, \qquad \qquad SEC^{\phi}_{3}= -\frac{8 m q^2_m}{5 \left(q^2_m+x^2\right)^{5/2}} \geq 0, \\\label{98}
DEC^{\phi}_{1}&=& \frac{8 m q^2_m}{5 \left(q^2_m+x^2\right)^{5/2}} \geq 0, \qquad \qquad DEC^{\phi}_{2}= -\frac{2q^2_m\left(5\sqrt{x^2+q^2_m}-14m\right)}{5(x^2+q^2_m)^{5/2}}  \geq 0,  \\\label{99}
DEC^{\phi}_{3}&=& WEC^{\phi}_{3}=  -\frac{q^2_m\left(5\sqrt{x^2+q^2_m}-14m\right)}{5(x^2+q^2_m)^{5/2}}   \geq 0.
\end{eqnarray}

\begin{eqnarray}\label{100}
NEC^{EM}_{2}&=& \frac{3 m q^2_m}{\left(q^2_m+x^2\right)^{5/2}}  \geq 0,  \qquad \qquad SEC^{EM}_{3}= \frac{18 m q^2_m}{5 \left(q^2_m+x^2\right)^{5/2}} \geq 0, \\\label{101}
DEC^{EM}_{1}&=& \frac{12 m q^2_m}{5 \left(q^2_m+x^2\right)^{5/2}} \geq 0,   \qquad \qquad DEC^{EM}_{2}= -\frac{3 m q^2_m}{5 \left(q^2_m+x^2\right)^{5/2}}  \geq 0,  \\\label{102}
DEC^{EM}_{3}&=& WEC^{EM}_{3}= \frac{6 m q^2_m}{5 \left(q^2_m+x^2\right)^{5/2}}\geq 0.
\end{eqnarray}

Analyzing the energy conditions for the scalar field outside the event horizon ($A>0$), we observe that the null energy condition $NEC^{\phi}_1$ Eq. (\ref{55}) is violated, which implies a violation of the dominant energy condition $DEC^{\phi}_1$. All other energy conditions for the scalar field are violated except for $NEC^{\phi}_2$ Eq. (\ref{56}). Regarding the energy conditions of the electromagnetic part, they are satisfied everywhere, except for the dominant energy condition $DEC^{EM}_2$ Eq. (\ref{101}).


The energy conditions for the region internal to the event horizon ($A<0$) explicitly for the model in question are defined by

\begin{eqnarray}\label{103}
NEC^{\phi}_{1}&=&NEC^{\phi}_{2}= \frac{2 q^2_m \left(\sqrt{q^2_m+x^2}-2 m\right)}{\left(q^2_m+x^2\right)^{5/2}} \geq 0, \qquad \qquad SEC^{\phi}_{3}= \frac{4q^2_m\left(5\sqrt{x^2+q^2_m}-12m\right)}{5(x^2+q^2_m)^{5/2}}\geq 0, \\\label{104}
DEC^{\phi}_{1}&=& DEC^{\phi}_{2}=  \frac{8 m q^2_m}{5 \left(q^2_m+x^2\right)^{5/2}}\geq 0,  \qquad \qquad DEC^{\phi}_{3}= WEC^{\phi}_{3}= \frac{q^2_m\left(5\sqrt{x^2+q^2_m}-6m\right)}{5(x^2+q^2_m)^{5/2}}   \geq 0.
\end{eqnarray}

Note that the electromagnetic energy conditions within the event horizon are the same as those defined for the outer region Eqs. (\ref{100}-\ref{102}).

Reexamining the energy conditions, we observe that the null energy conditions $NEC^{\phi}_1=NEC^{\phi}_2$, Eq. (\ref{103}), are violated within the event horizon. Specifically, focusing on the dominant energy conditions $DEC^{\phi}_1=DEC^{\phi}_2$, Eq. (\ref{104}), they are not violated, however, since these conditions include the null energy conditions, they are also violated. The strong energy condition $SEC^{\phi}_3$, Eq. (\ref{103}), is always violated within the horizon, as well as the dominant energy condition $DEC^{\phi}_3$, Eq. (\ref {104}).

\subsection{Bardeen-type model}\label{sec62}

Replacing the metric functions in the energy conditions, Eqs. (\ref{79}-\ref{84}), in the region where $A(x)>0$,  we have:

\begin{eqnarray}\label{105}
NEC^{\phi}_{1}&=& -\frac{2 q^2_m \left(\left(q^2_m+x^2\right)^{3/2}-2 m x^2\right)}{\left(q^2_m+x^2\right)^{7/2}} \geq 0, \qquad \qquad SEC^{\phi}_{3}=  \frac{8 m q^2_m \left(8 q^2_m-7 x^2\right)}{35 \left(q^2_m + x^2\right)^{7/2}} \geq 0, \\\label{106}
DEC^{\phi}_{1}&=& \frac{8 m q^2_m \left(7 x^2-8 q^2_m\right)}{35 \left(q^2_m +x^2\right)^{7/2}}  \geq 0, \qquad \qquad DEC^{\phi}_{2}= -\frac{2 q^2_m \left(32 m q^2_m-98 m x^2+35 \left(q^2_m +x^2\right)^{3/2}\right)}{35 \left(q^2_m+x^2\right)^{7/2}} \geq 0,  \\\label{107}
DEC^{\phi}_{3}&=& WEC^{\phi}_{3}=  -\frac{q^2_m \left(32 m q^2_m-98 m x^2+35 \left(q^2_m+ x^2\right)^{3/2}\right)}{35 \left(q^2_m +x^2\right)^{7/2}}   \geq 0.
\end{eqnarray}

For the electromagnetic part:

\begin{eqnarray}\label{108}
NEC^{EM}_{2}&=& \frac{m \left(13 q^2_m x^2-2 q^4_m\right)}{\left(q^2_m +x^2\right)^{7/2}}  \geq 0,  \qquad \qquad SEC^{EM}_{3}= \frac{6 m q^2_m \left(91 x^2-34 q^2_m\right)}{35 \left(q^2_m +x^2\right)^{7/2}} \geq 0, \\\label{109}
DEC^{EM}_{1}&=&  \frac{4 m q^2_m \left(16 q^2_m +91 x^2\right)}{35 \left(q^2_m +x^2\right)^{7/2}} \geq 0,   \qquad \qquad DEC^{EM}_{2}= \frac{m q^2_m \left(134 q^2_m -91 x^2\right)}{35 \left(q^2_m +x^2\right)^{7/2}} \geq 0,  \\\label{110}
DEC^{EM}_{3}&=& WEC^{EM}_{3}= \frac{2 m q^2_m \left(16 q^2_m +91 x^2\right)}{35 \left(q^2_m +x^2\right)^{7/2}} \geq 0.
\end{eqnarray}

The null energy condition for the scalar field $NEC^{\phi}_1$, Eq. (\ref{105}), is clearly violated outside the horizon since $A>0$. Therefore, the associated dominant energy condition $DEC^{\phi}_1$ Eq. (\ref{106}) is also violated. The secondary null energy condition $NEC^{\phi}_2$, Eq. (\ref{56}), is satisfied, while the dominant energy conditions $DEC^{\phi}_2$, Eq. (\ref{106}), and $DEC^{\phi}_3$, Eq. (\ref{107}), are violated. The strong energy condition $SEC^{\phi}_3$, Eq. (\ref{105}), is violated for throat radii between $- \sqrt{8/7}q_m>x>\sqrt{8/7}q_m$.

For the electromagnetic part, the main null energy condition $NEC^{EM}_1$ Eq. (\ref{61}) is satisfied within the horizon, as well as the dominant energy conditions $DEC^{EM}_1$ Eq. (\ref{109}) and $DEC^{EM}_3$ Eq. (\ref{110}). The secondary null energy condition $NEC^{EM}_2$ Eq. (\ref{108}) is violated for the throat radii between $- \sqrt{2/13}q_m<x<\sqrt{2/13}q_m$. Likewise, the dominant energy condition $DEC^{EM}_2$ Eq. (\ref{109}) is violated for the throat radii between $- \sqrt{134/91}q_m>x>\sqrt{134 /91}q_m$. Finally, the strong energy condition $SEC^{EM}_3$ Eq. (\ref{108}) is violated for the throat radii between $- \sqrt{34/91}q_m<x<\sqrt{34/ 91}q_m$.

The energy conditions for the region interior to the event horizon are given by

\begin{eqnarray}\label{111}
NEC^{\phi}_{1}&=& NEC^{\phi}_{2}= \frac{2q^2_m\left((x^2+q^2_m)^{3/2}-2mx^2\right)}{(x^2+q^2_m)^{7/2}} \geq 0, \\\label{112}
SEC^{\phi}_{3}&=& \frac{4 q^2_m \left(16 m q^2_m-84 m x^2+35 \left(q^2_m+x^2\right)^{3/2}\right)}{35 \left(q^2_m+x^2\right)^{7/2}} \geq 0, \\\label{113}
DEC^{\phi}_{1}&=& DEC^{\phi}_{2}= \frac{8 m q^2_m \left(7 x^2-8 q^2_m\right)}{35 \left(q^2_m+x^2\right)^{7/2}}\geq 0, \\\label{114}
DEC^{\phi}_{3}&=& WEC^{\phi}_{3}= \frac{q^2_m \left(-32 m q^2_m-42 m x^2+35 \left(q^2_m+x^2\right)^{3/2}\right)}{35 \left(q^2_m+x^2\right)^{7/2}}    \geq 0.
\end{eqnarray}

In the analysis of the energy conditions within the event horizon, $A<0$, clearly the null energy conditions $NEC^{\phi}_{1}= NEC^{\phi}_{2}$ Eq. (\ref{111}) are violated, therefore the dominant energy conditions are also violated $DEC^{\phi}_{1}= DEC^{\phi}_{2}$ Eq. (\ref{113}).
Finally, the strong energy condition $SEC^{\phi}_{3}$ Eq. (\ref{112}) is not violated within the horizon while the dominant energy condition $DEC^{\phi} _{3}$ Eq. (\ref{114}) is violated.

\section{Conclusion}\label{sec7}

In the present work, we start from an action that describes the k-essence theory with a scalar field, assuming a power form. 
This theory has been used to investigate black-bounce solutions \cite{CDJM1} and generalizations for different scalar field configurations \cite{CDJM2}. We extend this framework by introducing NED \cite{MM1} to explore possible charged black-bounce solutions for scalar field strengths that differ from the canonical case $n=1$. Specifically, we construct magnetic solutions such that the throat of the wormhole coincides with the magnetic charge $a=q_m$.

From the equations of motion, Eqs. (\ref{14}-\ref{17}), we can observe that the electromagnetic function $L_f$, contained in the equation Eq. (\ref{15}), does not depend on the form of the scalar field, but rather only on the metric functions $A(x)$ and $\Sigma(x)$. This leads to the conclusion that this function is the same as obtained for the canonical scalar field depending only on the chosen model.

We analytically derive all the functions involved for the Simpson-Visser model (Section \ref{sec3}) and for a Bardeen-type solution (Section \ref{sec4}) for k-essence configurations with $n=1/3$ and $n =1/5$. We can verify that for each model of a specific form, the electromagnetic functions $L_f(x)$ and $L(x)$ do not change with variation in the power of the k-essence field. In fact, these functions are the same as those obtained in the canonical case and investigated in \cite{INTRO8,MM1}. This behavior implies that the modifications in the scalar field due to k-essence are counterbalanced by the potential, keeping $L_f(x)$ and $L(x)$ unchanged. This behavior can be extended to the other powers of the phantom scalar field $n=1/3,1/5,1/7,\dots$.

We graphically represent the scalar field, potential, and electromagnetic functions for each of the models studied. Qualitatively, the electromagnetic functions $L_f(x)$ and $L(x)$ for both the Simpson-Visser model (Fig. \ref{SV1}) and the Bardeen-type model (Fig. \ref{BD1}) exhibit similar behavior. Regarding the behavior of the scalar field for both models in this configuration, it tends to invert its sign asymptotically $x\to{\pm\infty}$ for some charge values, as illustrated in Eqs. (\ref{19a}) and (\ref{32ASSIM}). The potential for both models and configurations tends to behave similar to a potential barrier for regions outside the horizon and creates minima as it becomes more internal to the horizon.

Finally, we analyzed the energy conditions for the scalar and electromagnetic fields for each of the models investigated. As previously observed in \cite{CDJM2}, the energy conditions for the scalar field do not change with the power of the k-essence, so it suffices to analyze only the case of the lower power $n=1/3$. Thus, in the Simpson-visser model, Section \ref{sec61}, the null energy condition $NEC^\phi_1$ and its associated dominant energy condition $DEC^\phi_1$ are violated outside the horizon, $A>0$. As well as all other energy conditions, except for $NEC^\phi_2$. For within the event horizon, $A<0$, all energy conditions are violated, Eqs. (\ref{103}-\ref{104}). Regarding the analysis of electromagnetic energy conditions for the Simpson-Visser model, all conditions are satisfied everywhere, except $DEC^{EM}_2$, Eq. (\ref{101}).

In the analysis of the energy conditions of the second model, Section \ref{sec62}, the energy conditions $NEC^\phi_1$, $DEC^\phi_1$, $DEC^\phi_2$, and $DEC^\phi_3$ are violated where $A>0$. The null energy condition $NEC^\phi_2$ is satisfied and $SEC^\phi_3$, Eq. (\ref{105}), is conditionally violated. Regarding the electromagnetic part, the energy conditions $NEC^{EM}_1$, $DEC^{EM}_1$, and $DEC^{EM}_3$ are satisfied, and the others are conditionally violated Eqs. (\ref {108}-\ref{110}). In regions where $A<0$, the null energy conditions $NEC^\phi_1$ and $NEC^\phi_2$ are violated, Eq. (\ref{111}), leading to the violation of the dominant energy conditions $DEC^{\phi}_1$, $DEC^{\phi }_2$. The energy density is violated, but $SEC^{\phi}_3$, Eq. (\ref{113}), is satisfied.

We intend to analyze the stability of these classes of black-bounces in future work. We hope that this work will begin to obtain new solutions consistent with some astrophysical data, such as shadows and gravitational waves.

\begin{acknowledgments}
We thank CNPq, CAPES, FAPES, and FUNCAP for financial support.
\end{acknowledgments}

\clearpage

\nocite{*}

\begin{thebibliography}{99}


\bibitem{INTRO1} R. D'Inverno, Introducing Einstein's Relativity, Oxford University Press, New York (1998).

\bibitem{INTRO2} S. Weinberg; Gravitation and Cosmology: principles and applications of the general theory of relativity, 1972.

\bibitem{INTRO3} M. P. Hobson, G. P. Efstathiou e A. N. Lasenby, General Relativity-An Introduction for Physicists, Cambridge University Press, Nova York (2006).

\bibitem{INTRO4} S. Chandrasekhar, The mathematical theory of black holes, Oxford University Press, Nova York (2006).

\bibitem{INTRO5} R. M. Wald, “General Relativity”, The University of Chicago Press, Chicago (1984).

\bibitem{LIGOScientific:2016aoc}
B.~P.~Abbott \textit{et al.} [LIGO Scientific and Virgo],
Phys. Rev. Lett. \textbf{116}, no.6, 061102 (2016),
[arXiv:1602.03837 [gr-qc]].

\bibitem{INTRO6} J. M. Bardeen, 
in Proceedings of the International Conference GR5, Tbilisi, U.S.S.R. (1968).

\bibitem{INTRO7} E. Ayon-Beato and A. Garcia, 
Phys. Lett. B \textbf{493}, 149-152 (2000), [arXiv:gr-qc/0009077 [gr-qc]].


\bibitem{Rodrigues:2018bdc}
M.~E.~Rodrigues and M.~V.~de Sousa Silva,
JCAP \textbf{06}, 025 (2018),
[arXiv:1802.05095 [gr-qc]].

\bibitem{Bronnikov:2022ofk}
K.~A.~Bronnikov,
[arXiv:2211.00743 [gr-qc]].


\bibitem{Bolokhov:2024sdy}
S.~V.~Bolokhov, K.~A.~Bronnikov and M.~V.~Skvortsova,
[arXiv:2405.09124 [gr-qc]].

\bibitem{Bronnikov:2024izh}
K.~A.~Bronnikov,
[arXiv:2404.14816 [gr-qc]].

\bibitem{Rodrigues:2017tfm}
M.~E.~Rodrigues and E.~L.~B.~Junior,
Phys. Rev. D \textbf{96}, no.12, 128502 (2017)
[arXiv:1712.03592 [gr-qc]].

\bibitem{matt} A.~Simpson and M.~Visser,
JCAP \textbf{02}, 042 (2019),
[arXiv:1812.07114 [gr-qc]].

\bibitem{INTRO8}  K.~A.~Bronnikov and R.~K.~Walia,
Phys. Rev. D \textbf{105}, no.4, 044039 (2022),
[arXiv:2112.13198 [gr-qc]].

\bibitem{INTRO9} P.~Ca\~nate,
Phys. Rev. D \textbf{106}, no.2, 024031 (2022),
[arXiv:2202.02303 [gr-qc]].

\bibitem{INTRO10} G.~Alencar, K.~A.~Bronnikov, M.~E.~Rodrigues, D.~S\'aez-Chill\'on G\'omez and M.~V.~d.~S.~Silva,
[arXiv:2403.12897 [gr-qc]].

\bibitem{INTRO11} M.~E.~Rodrigues and M.~V.~d.~Silva,
Class. Quant. Grav. \textbf{40}, no.22, 225011 (2023),
[arXiv:2204.11851 [gr-qc]].

\bibitem{MM2}  F.~S.~N.~Lobo, M.~E.~Rodrigues, M.~V.~de Sousa Silva, A.~Simpson and M.~Visser,
Phys. Rev. D \textbf{103}, no.8, 084052 (2021)
[arXiv:2009.12057 [gr-qc]].


\bibitem{RN1}E.~Franzin, S.~Liberati, J.~Mazza, A.~Simpson and M.~Visser,
JCAP \textbf{07}, 036 (2021),
[arXiv:2104.11376 [gr-qc]].


\bibitem{RN2} S.~Murodov, K.~Badalov, J.~Rayimbaev, B.~Ahmedov and Z.~Stuchl\'\i{}k,
Symmetry \textbf{16}, no.1, 109 (2024).

\bibitem{INTRO12}  E.~L.~B.~Junior and M.~E.~Rodrigues,
Gen. Rel. Grav. \textbf{55}, no.1, 8 (2023),
[arXiv:2203.03629 [gr-qc]].


\bibitem{INTRO13} J.~C.~Fabris, E.~L.~B.~Junior and M.~E.~Rodrigues,
Eur. Phys. J. C \textbf{83}, no.10, 884 (2023),
[arXiv:2310.00714 [gr-qc]].

\bibitem{INTRO14} J.~T.~S.~S.~Junior, F.~S.~N.~Lobo and M.~E.~Rodrigues,
Eur. Phys. J. C \textbf{84}, no.6, 557 (2024),
[arXiv:2405.09702 [gr-qc]].

\bibitem{INTRO15} K.~Atazadeh and H.~Hadi,
JCAP \textbf{01}, 067 (2024),
[arXiv:2311.07637 [gr-qc]].

\bibitem{INTRO16} T.~M.~Crispim, M.~Estrada, C.~R.~Muniz and G.~Alencar,
[arXiv:2405.08048 [hep-th]].

\bibitem{INTRO17} A.~Lima, G.~Alencar, R.~N.~Costa Filho and R.~R.~Landim,
Gen. Rel. Grav. \textbf{55}, no.10, 108 (2023),
[arXiv:2306.03029 [gr-qc]].

\bibitem{INTRO18} A.~Lima, G.~Alencar and D.~S\'aez-Chillon G\'omez,
Phys. Rev. D \textbf{109}, no.6, 064038 (2024),
[arXiv:2307.07404 [gr-qc]].

\bibitem{INTRO19} A.~M.~Lima, G.~M.~de Alencar Filho and J.~S.~Furtado Neto,
Symmetry \textbf{15}, no.1, 150 (2023),
[arXiv:2211.12349 [gr-qc]].


\bibitem{INTRO21} J.~Mazza, E.~Franzin and S.~Liberati,
JCAP \textbf{04}, 082 (2021),
[arXiv:2102.01105 [gr-qc]].



\bibitem{INTRO23} Z.~Xu and M.~Tang,
Eur. Phys. J. C \textbf{81}, no.10, 863 (2021),
[arXiv:2109.13813 [gr-qc]].

\bibitem{INTRO24}  J.~R.~Nascimento, A.~Y.~Petrov, P.~J.~Porfirio and A.~R.~Soares,
Phys. Rev. D \textbf{102}, no.4, 044021 (2020),
[arXiv:2005.13096 [gr-qc]].

\bibitem{INTRO25}N.~Tsukamoto,
Phys. Rev. D \textbf{104}, no.6, 064022 (2021),
[arXiv:2105.14336 [gr-qc]].

\bibitem{INTRO26} N.~Tsukamoto,
Phys. Rev. D \textbf{103}, no.2, 024033 (2021),
[arXiv:2011.03932 [gr-qc]].

\bibitem{INTRO27} S.~Ghosh and A.~Bhattacharyya,
JCAP \textbf{11}, 006 (2022),
[arXiv:2206.09954 [gr-qc]].
 
\bibitem{INTRO28} A.~Chowdhuri, S.~Ghosh and A.~Bhattacharyya,
Front. Phys. \textbf{11}, 1113909 (2023),
[arXiv:2303.02069 [gr-qc]].

\bibitem{INTRO29}  H.~Aounallah, A.~R.~Soares and R.~L.~L.~Vit\'oria,
Eur. Phys. J. C \textbf{80}, no.5, 447 (2020).

\bibitem{INTRO30} A.~R.~Soares, R.~L.~L.~Vit\'oria and H.~Aounallah,
Eur. Phys. J. Plus \textbf{136}, no.9, 966 (2021).

\bibitem{INTRO31}C.~F.~S.~Pereira, A.~R.~Soares, R.~L.~L.~Vit\'oria and H.~Belich,
Eur. Phys. J. C \textbf{83}, no.4, 270 (2023).

\bibitem{INTRO32}  C.~F.~S.~Pereira, R.~L.~L.~Vit\'oria, A.~R.~Soares and H.~Belich,
Mod. Phys. Lett. A \textbf{38}, no.28n29, 2350133 (2023).

\bibitem{INTRO33} J.~R.~Nascimento, G.~J.~Olmo, P.~J.~Porf\'\i{}rio, A.~Y.~Petrov and A.~R.~Soares,
Phys. Rev. D \textbf{101}, no.6, 064043 (2020),
[arXiv:1912.10779 [hep-th]].

\bibitem{INTRO34} M.~Barriola and A.~Vilenkin,
Phys. Rev. Lett. \textbf{63}, 341 (1989).

\bibitem{Rodrigues:2017yry}
M.~E.~Rodrigues, E.~L.~B.~Junior and M.~V.~de Sousa Silva,
JCAP \textbf{02}, 059 (2018),
[arXiv:1705.05744 [physics.gen-ph]].

\bibitem{PRL} K.~A.~Bronnikov and J.~C.~Fabris,
Phys. Rev. Lett. \textbf{96}, 251101 (2006),
[arXiv:gr-qc/0511109 [gr-qc]].


\bibitem{CDJM1}
C.~F.~S.~Pereira, D.~C.~Rodrigues, J.~C.~Fabris and M.~E.~Rodrigues,
Phys. Rev. D \textbf{109}, no.4, 044011 (2024),
[arXiv:2309.10963 [gr-qc]].

\bibitem{CDJM2} C.~F.~S.~Pereira,  D.~C.~Rodrigues, \'E.~L.~Martins, J.~C.~Fabris and M.~E.~Rodrigues, Class. Quant. Grav. \textbf{42} (1), 015001 (2025),[arXiv:2405.07455 [gr-qc]].

\bibitem{phan1} H.~G.~Ellis,
J. Math. Phys. \textbf{14}, 104-118 (1973).

\bibitem{phan2} K.~A.~Bronnikov,
Acta Phys. Polon. B \textbf{4}, 251-266 (1973).

\bibitem{phan3}  C.~R.~Almeida, J.~C.~Fabris, F.~Sbis\'a and Y.~Tavakoli,
[arXiv:1604.00624 [gr-qc]]. To appear in the proceedings of the 31st International Colloquium on Group Theoretical Methods in Physics.

\bibitem{phan4} K.~A.~Bronnikov, J.~C.~Fabris, O.~F.~Piattella, D.~C.~Rodrigues and E.~C.~Santos,
Eur. Phys. J. C \textbf{77}, no.6, 409 (2017),
[arXiv:1701.06662 [gr-qc]].

\bibitem{KDJ}  K.~A.~Bronnikov, J.~C.~Fabris and D.~C.~Rodrigues,
Grav. Cosmol. \textbf{22}, no.1, 26-31 (2016),
[arXiv:1511.08036 [gr-qc]].



\bibitem{MM1} M.~E.~Rodrigues and M.~V.~d.~S.~Silva,
Phys. Rev. D \textbf{107}, no.4, 044064 (2023),
[arXiv:2302.10772 [gr-qc]].


\bibitem{LZ}  M. Visser, Lorentzian Wormholes: from Einstein to Hawking, Springer-Verlag, Nova York (1996).
\end{thebibliography}
		
\end{document}